\documentclass[pra,twocolumn,showpacs,superscriptaddress,floatfix, nofootinbib]{revtex4-1}
\usepackage[caption=false]{subfig}
\usepackage{graphicx}
\graphicspath{ {} }
\usepackage{amsmath,graphicx}
\usepackage{epsfig}
\usepackage{amsmath,amssymb,mathrsfs,esint}
\usepackage{gensymb}
\usepackage[bookmarks=true,
   colorlinks=true,
   linkcolor=blue,
   urlcolor=blue,
   citecolor=blue,
   bookmarks=true,
   hyperindex=true
]{hyperref}
\usepackage{natbib}
\usepackage{relsize}
\usepackage{float}
\usepackage{dsfont}

\newcommand{\beq}{\begin{eqnarray}}
\newcommand{\eeq}{\end{eqnarray}}

\newcommand{\be}{\begin{equation}}
\newcommand{\ee}{\end{equation}}

\begin{document}

\title{One-dimensional two-component fermions with contact even-wave repulsion and SU(2) breaking near-resonant odd-wave attraction}

\author{D.\,V. Kurlov}

\affiliation{Van der Waals-Zeeman Institute, Institute of Physics, University of Amsterdam, Science Park 904, 1098 XH Amsterdam, The Netherlands}
\affiliation{LPTMS, CNRS, Univ. Paris-Sud, Universit\'e Paris-Saclay, Orsay 91405, France}

\author{S.\,I. Matveenko}
\affiliation{LPTMS, CNRS, Univ. Paris-Sud, Universit\'e Paris-Saclay, Orsay 91405, France}
\affiliation{L.D. Landau Institute for Theoretical Physics, Russian Academy of Sciences, Moscow 119334, Russia}

\author{V. Gritsev}
\affiliation{Institute for Theoretical Physics, University of Amsterdam, Science Park 904, 1098 XH Amsterdam, The Netherlands}
\affiliation{Russian Quantum Center, Skolkovo, Moscow 143025, Russia}

\author{G.\,V. Shlyapnikov}

\affiliation{Van der Waals-Zeeman Institute, Institute of Physics, University of Amsterdam, Science Park 904, 1098 XH Amsterdam, The Netherlands}
\affiliation{LPTMS, CNRS, Univ. Paris-Sud, Universit\'e Paris-Saclay, Orsay 91405, France}
\affiliation{Russian Quantum Center, Skolkovo, Moscow 143025, Russia}
\affiliation{SPEC, CEA, CNRS, Univ. Paris-Saclay, CEA Saclay, Gif sur Yvette 91191, France}

\begin{abstract}
We consider a one-dimensional (1D) two-component atomic Fermi gas with contact interaction in the even-wave channel (Yang-Gaudin model) and study the effect of an SU(2) symmetry breaking near-resonant odd-wave interaction within one of the components. Starting from the microscopic Hamiltonian, we derive an effective field theory for the spin degrees of freedom using the bosonization technique. It is shown that at a critical value of the odd-wave interaction there is a first-order phase transition from a phase with zero total spin and zero magnetization to the spin-segregated phase where the magnetization locally differs from zero.
\end{abstract}

\maketitle

\section{Introduction}

Over the last decades there has been a tremendous progress in the field of ultracold atomic quantum gases~\cite{Bloch2008, Guan2013, Massignan2014}. Unprecedented degree of precision, tunability, and control allows one to study an immense diversity of physical systems that are of great interest from the condensed matter physics perspective. Ultracold gases of atomic fermions that are in two internal states can be mapped onto spin-1/2 fermions treating the internal energy states as pseudospin states. These gases serve as an ideal platform for simulating a large variety of magnetically ordered phases, in particular, an itinerant ferromagnetic state.  

Itinerant ferromagnetism of spin-1/2 fermions in condensed matter systems is a long-standing problem \cite{Vollhardt2001_Lohneysen2007, Moriya1984_85, Tasaki1998}. It has also been intensively studied  in the context of ultracold quantum gases (see Ref. \cite{Massignan2014} for a review). It is believed that the Stoner criterion \cite{Stoner1933} (which requires a strong repulsion between different spin components) alone will not lead to the formation of an itinerant ferromagnetic state \cite{Pekker2011, Sanner2012}.
Despite a vast amount of experimental \cite{Jo2009, Sanner2012, Scazza2017, Valtolina2017, Amico2018} and theoretical \cite{Pekker2011, theor, Yang2004, Kozii2017} studies on itinerant ferromagnetism, many aspects, such as the character of the ferromagnetic phase transition remain disputable \cite{Massignan2014}. Recent experimental advances in time-resolved spectroscopic techniques \cite{Scazza2017, Amico2018} provide new prospects for studying itinerant ferromagnetism in ultracold quantum gases and renew the interest to this intriguing topic.

In the one-dimensional case, it has been realized long ago that the Stoner criterion is not valid. According to the Lieb-Mattis theorem \cite{Lieb1962, Aizenman1990}, in a one-dimensional two-component Fermi gas with a contact repulsive interaction between different spin species, the ferromagnetic state has a higher energy for any finite repulsion strength. In the limit of infinitely strong repulsion all spin configurations are degenerate \cite{Ogata1990, Shiba1991}. Recently, it has been shown that the itinerant ferromagnetic ground state can be realized in a 1D two-component Fermi gas with an infinite \cite{Jiang2016} or a very strong \cite{Yang2016} contact interspecies repulsion and an odd-wave attraction within one of the components. These proposals are very promising as they require regimes of the interactions that are reachable already with the present experimental facilities \cite{40K}.
However, the regime of finite and moderate repulsion strength has not been investigated. It is the purpose of this paper to fill in this gap.

We use bosonization and renormalization group (RG) techniques to study an effective field theory for a one-dimensional two-component Fermi gas with a contact repulsion between different components and an odd-wave attractive interaction within one of the components. The contact repulsive interaction is assumed to be in the weak or intermediate regime and the odd-wave attraction in the near resonant regime (precise definitions will be given below). It is shown that at a critical value of the odd-wave interaction there is a first-order phase transition from a phase with zero total spin and zero magnetization to the spin-segregated phase where the magnetization locally differs from zero.

This paper is organized as follows. In Section \ref{S:int} we specify the microscopic model and describe the interactions between the particles. In Section \ref{S:bos} we derive an effective field theory for the spin degrees of freedom using the bosonization technique. The resulting field theory is then studied using renormalization group analyzis in Section \ref{S:RG}, and in Section \ref{S:PT} we derive the phase transition criterion. Finally, in Section \ref{S:concl} we conclude. 

\section{Interaction between particles and Hamiltonian of the system} \label{S:int}

We begin with a brief description of the model. Consider a two-component one-dimensional atomic Fermi gas in free space at zero temperature. The total Hamiltonian is $H = H_0 + H_{\uparrow\downarrow} + H'$, which includes the free part $H_0$ and two types of the interaction. The interaction between different (pseudo)spin species, $H_{\uparrow\downarrow}$, is assumed to be contact and repulsive, and it takes place in the even-wave scattering channel. The term $H'$ describes the intraspecies attractive interaction in the odd-wave channel. Let us now discuss both interactions in detail.

\subsection{Even-wave interaction}
In the absence of the odd-wave interaction, the model reduces to the well-known Yang-Gaudin model with Hamiltonian $H_{\text{YG}} = H_0 + H_{\uparrow\downarrow}$, which explicitly reads (we use units in which $\hbar = 1$, unless specified otherwise):
\be \label{H_YG}
	 H_{\text{YG}} = \int dx \Bigl\{ -\frac{1}{2 m} \sum_{j = \uparrow,\downarrow} \psi^{\dag}_{j}\partial_x^2 \psi_{j} + g \psi^{\dag}_{\uparrow}\psi^{\dag}_{\downarrow}\psi_{\downarrow}\psi_{\uparrow} \Bigr\}.
\ee
Here $\psi_{j}$ is the field operator for a fermion in the (pseudo)spin state $j = \uparrow, \downarrow$ and $g$ is the even-wave interaction coupling constant. The model is exactly solvable \cite{Gaudin1967, Yang1967} and it is well known that for any finite repulsion ($g>0$) the ground state has total spin $S=0$. In the limit of infinite repulsion strength $g\to+\infty$ all spin configurations are degenerate \cite{Ogata1990}. Thus, in agreement with the Lieb-Mattis theorem, the even-wave contact repulsion alone cannot lead to the ground state with nonzero total spin \cite{Lieb1962, Aizenman1990}. The situation changes if one takes into account the interaction in the odd-wave scattering channel. This interaction is momentum-dependent and the Lieb-Mattis theorem no longer applies.


\subsection{Odd-wave interaction} \label{section_scatt_amp}
For ultracold fermions the background odd-wave interaction is rather weak, since it is proportional to the square of the relative momentum of colliding particles. Nevertheless, the interaction strenght can be enhanced using a Feshbach resonance. Just like $p$-wave interaction in higher dimensions, odd-wave interaction in one dimension takes place in the spin-triplet state of colliding particles. However, under realistic conditions a Feshbach resonance is usually only present for one of the states out of the triplet. For example, in the case of~$^{40}$K atoms there is a $p$-wave resonance at magnetic field 198.8\,G \cite{Jin2002, Regal2003}, and it is present only between two atoms in the $\left| F = 9/2, m_F = -7/2 \right>$ states. Therefore, in the presence of the Feshbach resonance the odd-wave interaction, typically, is not SU$(2)$-invariant.

The case of 1D spin-polarized fermions with resonant odd-wave interaction has been studied previously by means of the asymptotic Bethe ansatz \cite{Imambekov2010}. This approach relies on the knowledge of the two-body scattering phase shift (first derived in Ref. \cite{Pricoupenko2008}) and does not require an explicit form of the Hamiltonian. In the two-component case that we are dealing with, the asymptotic Bethe ansatz becomes cumbersome due to the existence of both charge and spin excitations. For this reason, we proceed differently and take into account the resonant odd-wave interaction using the so-called two-channel model that accurately captures microscopic physics of the interaction. This model describes the Feshbach resonant interaction as an interconversion between pairs of fermionic atoms in the open channel and weakly bound bosonic dimers in the closed channel \cite{sWave2ch, Gurarie2007, Prem2018}. 

Thus, keeping in mind the absence of SU(2) symmetry, we now include an odd-wave interaction within one of the components, say, between spin-$\uparrow$ fermions. The corresponding two-channel Hamiltonian in the momentum representation reads \cite{Cui2016}
\be \label{H_odd}
\begin{aligned}
	 & H' = \sum_q \left( \frac{q^2}{4 m} + \nu \right) b^{\dag}_q  b_q \\
	 &+ \frac{\lambda}{\sqrt{L}} \sum_{k_1, k_2}  \frac{k_1 - k_2}{2} \left[ b^{\dag}_{k_1+k_2} a_{k_1, \uparrow} a_{k_2, \uparrow} + \text{H.c.} \right].
\end{aligned}
\ee
Here $\hat a^{\dag}_{k,\uparrow}$ is the fermionic creation operator of an (open channel) atom in the spin-$\uparrow$ state with mass $m$ and momentum $k$. The bosonic operator $\hat b^{\dag}_q$ creates an odd-wave (closed channel) dimer of spin-$\uparrow$ atoms with mass $2m$ and a center of mass momentum $q$. We denote by $\lambda$ the atom-dimer interconversion strength and the bare detuning of a dimer by $\nu$.
The latter is related to the dimer binding energy and can be tuned by an external magnetic field. 
The odd-wave interaction is momentum-dependent, and we introduce an ultraviolet momentum cutoff $\Lambda$, above which the interconversion strength $\lambda$ vanishes. 

One can relate the bare parameters of the odd-wave interaction ($\lambda$ and $\nu$) to the physical scattering parameters by calculating diagrammatically the two-body scattering amplitude \cite{Cui2016}:
\be \label{f_odd}
	f(k) = \frac{-i k}{- \hbar^2 \nu/m \lambda^2 + 2\Lambda/\pi + (\hbar^4/m^2 \lambda^2)k^2 + i k},
\ee
where we restored $\hbar$ for clarity. Comparing Eq. (\ref{f_odd}) with the general form of the 1D odd-wave scattering amplitude at low collisional energy, $f(k) = -i k / [1/l_p + \xi_p k^2 +i k ]$, where $l_p$ is the 1D odd-wave scattering length and $\xi_p$ is the 1D effective range \cite{Pricoupenko2008}, we find:
\be \label{l_p_xi_p}
	l_p = -\,\frac{m \lambda^2}{\hbar^2} \frac{1}{\nu - 2 m\lambda^2 \Lambda / \pi \hbar^2}, \qquad \xi_p = \frac{\hbar^4}{m^2 \lambda^2}.
\ee
We see that the momentum cutoff $\Lambda$ simply results in the renormalization of the bare detuning $\nu$, similarly to the case of $s$-wave Feshbach resonant scattering in 3D \cite{Gurarie2007, Levinsen2011}.

Attractive odd-wave interaction corresponds to $l_p < 0$.
It follows from Eq. (\ref{l_p_xi_p}) that in this case the bare detuning is necessarily positive and satisfies the condition
\be \label{nu_attraction}
	\nu > \frac{2\Lambda}{\pi} \frac{\hbar^2}{m \xi_p}.
\ee
Let us estimate the right hand side of inequality (\ref{nu_attraction}). In the quasi-1D regime obtained by a tight harmonic confinement in transverse directions with frequency $\omega_{\perp}$ the 1D effective range can be written as $\xi_p = \alpha_1 a_{\perp}^2/3$, where $a_{\perp} = \sqrt{\hbar/m\omega_{\perp}}$ is the oscillator length and $\alpha_1$ is the 3D effective range \cite{Pricoupenko2008}. 
Eq. (\ref{nu_attraction}) then becomes $\nu > (6 \Lambda/\pi \alpha_1) \hbar \omega_{\perp}$, where the right hand side can now be easily estimated. 
Indeed, the 1D regime requires that $\hbar \omega_{\perp} \gg E_F$ and $a_{\perp} \gg R_e$, where $E_F$ is the Fermi energy and $R_e$ is the effective radius of the actual interaction potential between atoms.
The first condition gives $1/a_{\perp} \gg k_F$, and hence we may put the momentum cutoff to be $\Lambda \sim 1/a_{\perp}$. Then, the second condition implies that the ratio $\Lambda/\alpha_1 \ll 1$, since the 3D effective range $\alpha_1$ is typically of the order of $R_e^{-1}$. 
Therefore, for attractive interactions the lower bound of the bare detuning $\nu$ is of the order of $E_F$. As $\nu$ approaches this lower bound one enters the regime of resonant interactions, where $|l_p|$ is very large. On the contrary, in the off-resonant regime, where $|l_p|$ is small, the bare detuning is $\nu \gg E_F$. Then, from Eq. (\ref{l_p_xi_p}) we have $l_p \approx - m \lambda^2/ \hbar^2 \nu$.


\subsection{Effective odd-wave interaction}
In this subsection we integrate out the closed channel bosonic dimers and obtain an effective action for the fermionic fields.
The Euclidean action corresponding to Hamiltonian (\ref{H_odd}) is 
\be \label{S_odd}
	S' = \int d\tau dx \left\{ \bar \chi \left( \partial_\tau  - \frac{1}{4m} \partial^2_x + \nu \right) \chi +\lambda \Bigl( \bar\chi {\cal O} + \bar{\cal O} \chi \Bigr) \right\},
\ee
where $\tau = i t $ is the imaginary time, $\chi$ and $\bar \chi$ are bosonic complex fields, and we defined 
\be \label{O_op}
	{\cal O}(x,\tau) = \psi_{\uparrow} \bigl(i \partial_x \psi_{\uparrow}\bigr) - \bigl(i \partial_x \psi_{\uparrow}\bigr) \psi_{\uparrow}.
\ee
Integrating out the bosonic fields we obtain an effective action that contains only the fermionic fields:
\be \label{S_odd_eff}
	S'_{\text{eff}}= \lambda^2 \int d{\bf 1} d{\bf 2} \; \bar{\cal O} ({\bf 1}) {\cal G}({\bf 1}-{\bf 2}) {\cal O}({\bf 2}),
\ee
where ${\bf 1} \equiv (x, \tau)$ and ${\bf 2} \equiv (x', \tau')$. The bosonic propagator ${\cal G}(x,\tau)$ satisfies the equation
\be
	\left( -\partial_\tau + \frac{1}{4m} \partial^2_x -\nu \right) {\cal G}(x,\tau) = \delta(x)\delta(\tau)
\ee
and at zero temperature it reads
\be \label{greens_fun}
	{\cal G}(x,\tau) = - \theta(\tau) \sqrt{\frac{m}{\pi \tau}} \exp\left\{-\frac{m x^2}{\tau} - \nu \tau\right\}.
\ee
We see that, since $\nu$ is large and positive, the propagator is strongly localized in the vicinity of $x = \tau = 0$.

\section{Bosonization Procedure} \label{S:bos}
\subsection{Notations}
We now focus on the low-energy scattering near the Fermi points. In this subsection we briefly discuss the notations that we are going to use. The fermionic field is decomposed as
\be \label{RL_decompose}
	\psi_{j}(x, \tau) \approx e^{i k_F x} R_{j}(x, \tau) + e^{-i k_F x} L_{j}(x, \tau),
\ee
where $j = \uparrow,\downarrow$ is the spin index and $k_F = \pi n /2$ is the single-component Fermi momentum, with $n$ being the total fermionic density. For the slow fields $R_{j}$, $L_{j}$ we employ the bosonization identity in the following form:
\be \label{RL_bos_identity}
\begin{aligned}
	R_{j}(x, \tau) &= \frac{1}{\sqrt{2\pi a}} \, e^{-i \sqrt{\pi} \left\{ \Phi_{j}(x,\tau) + \Theta_{j}(x,\tau) \right\}},\\
	L_{j}(x,\tau) &= \frac{1}{\sqrt{2\pi a}} e^{i \sqrt{\pi} \left\{ \Phi_{j}(x,\tau) - \Theta_{j}(x,\tau) \right\}},
\end{aligned}
\ee
where $a$ is the short distance cut-off.
In equation (\ref{RL_bos_identity}), $\Phi_{j}$ is the compact bosonic field and $\Theta_{j}$ is the corresponding dual field. The latter is defined as
\be
	 \Theta_{j}(x,\tau) = - \int_{-\infty}^x dy \, \Pi_{j}(y,\tau),
\ee
where $\Pi_{j}$ is the canonical momentum conjugated to the field $\Phi_{j}$. Thus, one has the following equal time commutation relations $\left[ \Phi_j(x,\tau), - \partial_{x'}\Theta_j(x', \tau) \right] = i \delta(x-x')$.
For later purposes we also introduce a canonical momentum $\pi_j$ conjugated to the dual field $\Theta_j$. It is defined as
\be
	\pi_j(x,\tau) = - \int_{-\infty}^x dy \, \Theta_{j}(y,\tau)
\ee
and satisfies the commutation relations $\left[ \Theta_j(x,\tau), - \partial_{x'}\Phi_j(x', \tau) \right] = i \delta(x-x')$.
This will be useful once we turn to bosonizing the odd-wave interaction, as the latter acquires a very compact form in the dual representation.

\subsection{Even-wave interaction}
We now proceed with constructing a low-energy theory for our model.
The bosonized form of Eq. (\ref{H_YG}) is well known (see, e.g., Refs. \cite{Recati2003, Matveenko1994}):
\be \label{H_YG_bosonized}
\begin{aligned}
	 H_{\text{YG}} &= \sum_{\alpha = \rho,\sigma} \frac{u_{\alpha}}{2} \int dx \left\{  K_{\alpha} \Theta_{\alpha}'{}^2 + \frac{1}{K_{\alpha}} \Phi_{\alpha}'{}^2 \right\} \\
	&+ \frac{2 g}{(2\pi a)^2} \int dx \cos \sqrt{8\pi}\Phi_{\sigma} + H_{(3)},
\end{aligned}
\ee
where the charge and spin bosonic fields are $\Phi_{\rho,\sigma} = ( \Phi_{\uparrow} \pm \Phi_{\downarrow} )/\sqrt{2}$ and similarly for $\Theta_{\rho,\sigma}$. In order to lighten our notations, we denoted the spacial derivative by a prime. The charge (spin) velocity is $u_{\rho(\sigma)}$, and $K_{\rho(\sigma)}$ is the corresponding Luttinger parameter. In the regime of weak ($\gamma \ll 1$) and intermediate ($\gamma \lesssim 1$) repulsion strength they are given by
\be \label{Luttinger_params}
\begin{aligned}
	u_{\rho} &= v_F \left(1+ \frac{2\gamma}{\pi^2}\right)^{1/2}, \quad K_{\rho} = \left(1+ \frac{2\gamma}{\pi^2}\right)^{-1/2}, \\
	u_{\sigma} &= v_F \left(1- \frac{2\gamma}{\pi^2}\right)^{1/2}, \quad K_{\sigma} = \left(1- \frac{2\gamma}{\pi^2}\right)^{-1/2},
\end{aligned}
\ee
where $\gamma = m g/ n = \pi g/ 2 v_F$ is the dimensionless coupling constant, with $v_F$ being the Fermi velocity.
For repulsive interactions the cosine term in Eq. (\ref{H_YG_bosonized}) is irrelevant in the renormalization group sence and we omit it \cite{cosine}. The term $H_{(3)}$ comes from the spectrum nonlinearity in the vicinity of the Fermi points and reads \cite{ Matveenko1994, Haldane1981}:
\be
\begin{aligned}
	H_{(3)} = -\frac{\Gamma}{6} \int dx \Bigl\{ & \Phi'{}^3_{\rho} + 3 \Phi{}'_{\rho} \left( \Theta'{}^2_{\rho} + \Theta'{}^2_{\sigma} + \Phi'{}^2_{\sigma} \right) \\
	&+ 6 \, \Theta'_{\rho} \Theta'_{\sigma} \Phi'_{\sigma} \Bigr\},
\end{aligned}
\ee
where $\Gamma = \sqrt{\pi/2 m^2}$.

For later purposes we will need the Euclidean action written in terms of the $\Theta_{\rho, \sigma}$ fields only. Omitting the cosine term in Eq. (\ref{H_YG_bosonized}), we first write the dual field representation of the Hamiltonian density:
\be
\begin{aligned}
	&{\cal H}_{\text{YG}} = \sum_{j=\rho,\sigma} \left\{ \frac{u_j}{2 K_j} \pi_j^2 + \frac{u_j K_j}{2} \Theta_j'^2 \right\} \\
	&+ \frac{\Gamma}{6} \left\{ \pi_{\rho}^3 + 3\pi_{\rho} \left( \Theta_{\rho}'{}^2 + \Theta_{\sigma}'{}^2 + \pi_{\sigma}^2 \right) + 6 \, \Theta_{\rho}' \Theta_{\sigma}' \pi_{\sigma} \right\}.
\end{aligned}
\ee
Then, the corresponding Euclidean Lagrangian is ${\cal L}_{\text{YG}} = {\cal H}_{\text{YG}} - \sum_{j=\rho,\sigma} i \dot\Theta_{j} \pi_j$, where dot denotes the imaginary time derivative, and $\pi_{\rho,\sigma} = \pi_{\rho,\sigma} ( \dot\Theta_{\rho}, \dot\Theta_{\sigma} )$. The latter relation is obtained from $\pi_j = \partial {\cal H}_{\text{YG}} / \partial \dot \Theta_j$, which yields a set of nonlinear equations. Solving these equations for $\pi_j$ in terms of  $\dot\Theta_j$ to second order in $\Gamma$, we obtain:
\begin{widetext}
\be \label{L_YG}
	{\cal L}_{\text{YG}}  \approx \sum_{j=\rho,\sigma} \frac{K_j}{2} \left\{ \frac{1}{u_j} \dot\Theta_j^2  +  u_j \Theta_j '^{2} \right\} 
	+ \frac{\Gamma K_{\rho}}{2 u_{\rho} } \, i \, \dot\Theta_{\rho} \left\{ \Theta_{\sigma}'^2 - \frac{K_{\sigma}^2}{u_{\sigma}^2} \dot \Theta_{\sigma}^2 \right\} 
	- \frac{\Gamma ^2K_{\rho}}{8u_{\rho}} \left\{\Theta_{\sigma }'{}^4 +  \frac{K_{\sigma
   }^4}{u_{\sigma }^4} \dot\Theta _{\sigma}{}^4 \right\},
\ee
\end{widetext}
where we kept only the most relevant terms.

\subsection{Odd-wave interaction}
We now turn to bosonizing the effective action for the odd-wave interaction.
Substituting Eq. (\ref{RL_decompose}) into Eq. (\ref{O_op}) and keeping only the dominant term, we obtain
\be
	{\cal O}(x,\tau) \approx 2k_F R_{\uparrow}(x,\tau) L_{\uparrow}(x,\tau) = \frac{k_F}{\pi a} e^{- i \sqrt{4\pi} \Theta_{\uparrow} (x,\tau)}.
\ee
Other terms contain spatial derivatives of $R_{\uparrow}$ and $L_{\uparrow}$. For collisions taking place near the Fermi points such terms are suppressed since they are proportional to a small relative momentum of colliding particles. This can be easily seen, e.g., by using the decomposition (\ref{RL_decompose}) in Eq. (\ref{H_odd}). Therefore, we omit these terms and the effective action (\ref{S_odd_eff}) takes the following simple form:
\be \label{S_eff_final}
	S'_{\text{eff}} = \left(\frac{\lambda k_F}{\pi a} \right)^2 \int d{\bf 1} d{\bf 2} \;  {\cal G}({\bf 1}-{\bf 2}) \, e^{i \sqrt{4\pi} \Theta_{\uparrow}({\bf 1})} e^{-i \sqrt{4\pi} \Theta_{\uparrow}({\bf 2})}.
\ee
We then take into account that vertex operators multiply according to
\be \label{BCH}
	e^A e^B = \,:e^{A+B}:\, e^{\left\langle AB + \frac{A^2 + B^2}{2} \right\rangle_0},
\ee
where $: \ldots :$ denotes normal ordering and $\langle \ldots \rangle_0$ is the Gaussian average with respect to the bosonized Hamiltonian of free fermions. For the $\Theta_{\uparrow}$ fields one has
\be \label{corr_ThetaTheta}
\langle \Theta_{\uparrow}(x,\tau) \Theta_{\uparrow}(0,0) \rangle_0 - \langle \Theta_{\uparrow}^2(0,0) \rangle_0 = \frac{-1}{4\pi} \ln \frac{x^2 + v_F^2\tau^2 + a^2}{a^2}.
\ee
Note that we treat the even- and odd-wave interactions on equal footing and consider both of them as perturbations on top of free fermions. Therefore, in Eq.~(\ref{corr_ThetaTheta}) the Luttinger parameter is unity.
Thus, the effective action (\ref{S_eff_final}) becomes
\be \label{S_eff_bos2}
\begin{aligned}
	S'_{\text{eff}} &=  \left(\frac{\lambda k_F}{\pi} \right)^2 \int d{\cal T} d{\cal R} dt dr \, \frac{{\cal G}(r, t)}{r^2 + v_F^2 t^2 + a^2} \\
	 	&\times: e^{ i\sqrt{4\pi} \left[ \Theta_{\uparrow}\left({\cal R}+\frac{r}{2}, {\cal T}+\frac{t}{2}\right) - \Theta_{\uparrow}\left({\cal R}-\frac{r}{2},{\cal T}-\frac{t}{2}\right) \right] }: \, ,
\end{aligned}
\ee
where we switched to the coordinates $r = x-x'$, ${\cal R} = (x+x')/2$,  $t = \tau - \tau'$, and ${\cal T} = (\tau + \tau')/2$.
Note that the cut-off $a$ was cancelled. Due to the fact that ${\cal G}(r,t)$ differs from zero only for fairly small $r$ and $t$ [see Eq. (\ref{greens_fun})], we can expand the argument of the normal-ordered exponent in Taylor series.

Let us now discuss some general properties of the action (\ref{S_eff_bos2}). Expansion of the $\Theta_{\uparrow}$ fields and that of the normal-ordered exponent will produce a number of terms. To any order, any given term will obviously be a monomial in the ${\cal R}$ and ${\cal T}$ derivatives of $\Theta_{\uparrow}({\cal R},{\cal T})$. Its coefficient is a numerical constant times a monomial in $t$ and $r$. The latter contributes to the integral over $dr dt$. Therefore, it is convenient to define
\be \label{Isp}
\begin{aligned}
	&{\cal I}_{s,p} = \int_{\mathbb{R}^2} dr dt \, \frac{ r^{s} \, t^{p} \, {\cal G}(r, t)}{r^2 + v_F^2 t^2 + a^2} \\
	&= -\frac{1+(-1)^s}{\sqrt{\pi}} \frac{v_F^{s-2}}{(m v_F^2)^{s+p-1}} \int_0^{+\infty} \frac{dxdy\, x^s y^{p-\frac{1}{2}} }{x^2 + y^2 +a^2}  \,e^{-\frac{x^2}{y} - \tilde \nu y } ,
\end{aligned}
\ee
where $s$, $p$ are non-negative integers (not equal to zero simultaneously), and $\tilde \nu = \nu/(m v_F^2) = 1/(\eta \kappa)$, with dimensionless parameters
\be
\eta = k_F |l_p| \quad \text{and} \quad \kappa = k_F \xi_p.
\ee
The integral in Eq. (\ref{Isp}) can be calculated exactly. It converges for $\eta \kappa < 1$ and does not depend on the cutoff~$a$. We leave the details to Appendix \ref{A:integrals}. At this point it is only important to observe that ${\cal I}_{s,p} \leq 0$ for any $s$ and $p$. In other words, the overall coefficient in the action (\ref{S_eff_bos2}) is negative. For later convenience we define
\be \label{alpha_s_p}
\begin{aligned}
	&\left(\frac{\lambda k_F}{\pi} \right)^2 {\cal I}_{s,p} \equiv - \, \frac{v_F^{s-1}}{(m v_F^2)^{s+p-2}} \alpha_{s,p},\\
	& \text{where} \quad \alpha_{s,p} = \frac{1+(-1)^s}{\pi^{5/2} \kappa}  \tilde {\cal I}_{s,p} (\eta \kappa) \geq 0.
\end{aligned}
\ee
In Eq. (\ref{alpha_s_p}), $\tilde{\cal I}_{s,p}$ is the integral over $dx dy$ from Eq. (\ref{Isp}), and we took into account that $\lambda = 1/\sqrt{m^2 \xi_p}$, according to Eq. (\ref{l_p_xi_p}). For $\eta\kappa \ll 1$, i.e. not too close to the resonance, we have
\be \label{alpha_s_p2}
	\alpha_{s,p} \approx \frac{1+(-1)^s}{2 \pi^{5/2}} \Gamma\left( \frac{s+1}{2} \right) \Gamma\left( \frac{s-1}{2} \right) \frac{(\eta\kappa)^{p+\frac{s}{2}}}{\kappa},
\ee
for $s\neq 0$. In the case $s=0$, the corresponding expression for $\alpha_{s,p}$ is
\be
	\alpha_{0,p} \approx \frac{1}{\pi^{3/2}} \frac{(\eta \kappa)^{p - \frac{1}{2}}}{\kappa}.
\ee 

Written in terms of the spin and charge fields, the normal ordered exponent in Eq. (\ref{S_eff_bos2}) becomes
\be \label{norm_ord_exp_spin_charge}
	: e^{ i\sqrt{2\pi} \sum_{j={\rho,\sigma}} \left[ \Theta_{j}\left({\cal R}+\frac{r}{2}, T+\frac{t}{2}\right) - \Theta_{j}\left({\cal R}-\frac{r}{2},T-\frac{t}{2}\right) \right] }: \,.
\ee
It is now straightforward to write down the contribution from the odd-wave interaction to any desired order. Expanding the fields and the exponentials, we keep only the most relevant terms. These are the terms up to the second order in $\Theta_{\rho}$ (including terms like $\dot\Theta_{\rho} \Theta_{\sigma}'{}^2$) and to the sixth order in $\Theta_{\sigma}$. For the sake of readability in the main text we do not present them, but their explicit form is given in Appendix \ref{A:bosonization}. There we also provide more detailed calculations for the next section, where we obtain the total bosonized action.

\subsection{Total bosonized action}

Combining Eqs. (\ref{L_YG}) and (\ref{Lcharge}) - (\ref{Lspincharge}), we arrive at the Euclidean action $S = \int dx d\tau {\cal L}$, where the total Lagrangian is ${\cal L} = {\cal L}_{\sigma} + {\cal L}_{\rho}^{(0)} + {\cal L}_{\rho\sigma}$ with
\be \label{L_sigma}
\begin{aligned}
	&{\cal L}_{\sigma} = - i \varepsilon_{\sigma} \dot \Theta_{\sigma} + \frac{A_2^{\sigma} }{2} \dot\Theta_{\sigma}^2 + \frac{B_2^{\sigma}}{2} \Theta_{\sigma}'^2
	+ A_3^{\sigma}\, i\, \dot\Theta_{\sigma}^3 \\
	&-A_4^{\sigma} \dot\Theta _{\sigma}^4 
	- B_4^{\sigma} \Theta_{\sigma }'{}^4 
	-A_5^{\sigma} \,i\,\dot\Theta_{\sigma}^5
	+ A_6^{\sigma} \dot \Theta_{\sigma}^6
	+B_6^{\sigma} \Theta_{\sigma}'{}^6,
\end{aligned}
\ee
\be \label{L_rho}
	{\cal L}_{\rho}^{(0)} = - i \varepsilon_{\rho} \dot \Theta_{\rho} + \frac{A_2^{\rho}}{2} \dot\Theta_{\rho}^2 + \frac{B_2^{\rho}}{2} \Theta_{\rho}'^2,
\ee
\be \label{L_rho_sigma}
	{\cal L}_{\rho\sigma} = C\,  \Theta'_{\rho} \Theta'_{\sigma} 
	+ D\, \dot\Theta_{\rho} \dot\Theta_{\sigma} 
	+ i \, \dot\Theta_{\rho} \left( E  \,  \Theta_{\sigma}'^2 
	- F\,  \dot \Theta_{\sigma}^2 \right). 
\ee
The coefficients can be expressed in terms of $u_{\rho,\sigma}$, $K_{\rho,\sigma}$, $\Gamma$, and $\alpha_{s,p}$. For our purpose their explicit form is unimportant, but it can be found in Appendix \ref{spin_fields_eff_action}. 

Linear terms in Eqs. (\ref{Lspin}) and (\ref{Lcharge}) are allowed by symmetry and deserve comments. These are the so-called topological $\theta$-terms. It is well known that in a quantum problem such terms can play an important role. However, we are dealing with the zero temperature case, in which the quantum (1+1)-dimensional system under consideration is formally equivalent to a 2-dimensional classical field theory. For this reason we expect that in our case these topological terms do not lead to any subtle effects. Therefore, in the charge sector we take the linear term into account by simply shifting the field as follows:
\be \label{shift}
	\Theta_{\rho} = \tilde\Theta_\rho + i \beta_{\rho} \tau, \qquad \beta_{\rho} =  \varepsilon_{\rho}  /  A_2^{\rho} 
\ee

This shift does not influence the commutation relations and the integration measure in the partition function. Its only effect is to bring ${\cal L}_{\rho}^{(0)}$ to the form $(A_2^{\rho}/2) \dot{\tilde\Theta}_{\rho}^2 + (B_2^{\rho}/2) \tilde\Theta_{\rho}'^2$. After that the charge fields can be integrated out using standard methods. This results in the effective Lagrangian for the dual spin field $\Theta_{\sigma}$. We then rewrite it in terms of the spin density $\Phi_{\sigma}$ (for details see Appendix~\ref{spin_fields_eff_action}), and obtain the effective Lagrangian
\be \label{Lagr}
	\tilde{\cal L} ( \dot\Phi_{\sigma}, \Phi_{\sigma}' ) = \frac{A}{2} \dot\Phi_{\sigma}^2 + \frac{u_2}{2}\Phi_{\sigma}'^2 + \sum_{n=3}^6 u_n \Phi_{\sigma}'^n,
\ee
where for the coefficients to the leading order in $\eta \kappa \ll 1$ one has
\be \label{GL_coeffs}
\begin{aligned}
	A  &= \frac{1}{u_{\sigma }K_{\sigma }} \left( 1 +  \frac{y}{2 K_{\sigma}^2} \right), \;\; u_2  = \frac{u_{\sigma }}{K_{\sigma }} \left(1 -\frac{y}{2} \right), \\
	u_3 &= \sqrt{\frac{\pi}{32}} \left(\frac{u_{\sigma}}{K_{\sigma}}\right)^3 \frac{(\eta \kappa)^{5/2}}{m v_F^3\kappa}, \\
	u_4 &= - \frac{\pi}{32 m^2} \frac{K_{\rho}}{u_{\rho}} \left( 1 - 2 y \right), \\
	u_5 &= -\frac{3 \pi^{3/2}}{32 \sqrt{2} } \frac{K_{\rho}}{u_{\rho}} \left( \frac{u_{\sigma}}{K_{\sigma}} \right)^2 \frac{(\eta \kappa)^{5/2}}{m^3 v_F^3 \kappa}, \\
	u_6 &= \frac{\pi^2}{128} \frac{1}{m^4} \left(  \frac{K_{\rho}}{u_{\rho}} \right)^2 \frac{K_{\sigma}}{u_{\sigma}} \left( 1 - \frac{7}{2} y \right).
\end{aligned}
\ee
The dimensionless parameter $y$ is defined as
\be \label{y_def}
	y = K_{\sigma} \frac{v_F}{u_{\sigma}} \frac{\sqrt{\eta \kappa}}{\kappa} = \sqrt{\frac{\eta}{\kappa}} \, \left( 1 - \frac{2\gamma}{\pi^2} \right)^{-1}.
\ee
As can be seen from Eq. (\ref{GL_coeffs}), for $y<2/7 \approx 0.286$ we have $u_2 >0$, $u_4 < 0$, and $u_6 > 0$ \cite{instability}. 
From Eqs. (\ref{Luttinger_params}) and (\ref{y_def})  we find that in order to be in this regime one should have
\be
	\eta < \frac{4 \kappa}{49} \left( 1 - \frac{2\gamma}{\pi^2} \right)^2.
\ee
This condition can be easily satisfied.
Thus, already at the mean field level, the transition is of the first order.

\section{Renormalization Group Analysis} \label{S:RG}
Using Lagrangian (\ref{Lagr}), we write the Ginzburg-Landau functional
\be \label{GL_RG}
		\tilde S = \int d\tau d^dx \left\{ \frac{A}{2} \dot\Phi_{\sigma}^2 + \frac{u_2}{2} \Phi_{\sigma}'{}^2 + \sum_{n=3}^6 u_n \Phi_{\sigma}'{}^n \right\},
\ee 
with the coefficients given by Eq. (\ref{GL_coeffs}). For the RG analysis we have generalized the theory to the $d$-dimensional space. In order to gain insight into the role of the nonlinear terms, let us first look at their scaling behaviour. By changing $x = b \tilde x$, $\tau = b^z \tilde \tau$, and $\Phi_{\sigma}(x,\tau) = b^{\chi} \tilde \Phi_{\sigma}(\tilde x, \tilde \tau)$ we ontain
\be \label{tree_level_RG}
\begin{aligned}
	A(b) &= b^{2\chi + d - z}A, \qquad u_2 (b) = b^{2\chi + d + z - 2} u_2, \\
	u_n(b) & = b^{n\chi +d + z - n}u_n, \quad n = 3,\ldots,6.
\end{aligned}
\ee
Out of these quantities one can construct the following dimensionless couplings that are independent of the arbitrary rescaling exponents $z$ and $\chi$:
\be \label{g_n}
\begin{aligned}
	g_3 &= \frac{\Lambda \, u_3}{(\pi^2 A \, u_2^5 )^{1/4}}, \qquad g_4 = \frac{ \Lambda^2 \, u_4}{\pi \,( A\, u_2^3 )^{1/2}}, \\
	g_5 & = \frac{ \Lambda^3\, u_5}{(\pi^2 A^3\, u_2^7 )^{1/4}}, \qquad g_6 = \frac{\Lambda^4 \, u_6}{\pi\, A \,u_2^2}.
\end{aligned}
\ee
Then, using Eqs. (\ref{tree_level_RG}) and putting $b \approx 1 + \delta l$, one immediately finds that at the tree level the above nonlinear couplings flow according to the RG equations
\be
\begin{aligned}
	\partial_l g_3 &= - \frac{1}{2}(1+d) g_3, \qquad \partial_l g_4 = -(1+d)g_4,\\
	\partial_l g_5 &= - \frac{3}{2}(1+d) g_3, \qquad \partial_l g_6 = -2 (1+d)g_6.
\end{aligned}
\ee
Let us note that at this level the combination $g_6/g_4^2 = \pi u_2 u_6/u_4^2$ is invariant under the RG flow. 

\begin{widetext}

\begin{figure}[t]
\subfloat{%
  \includegraphics[width=.4\linewidth]{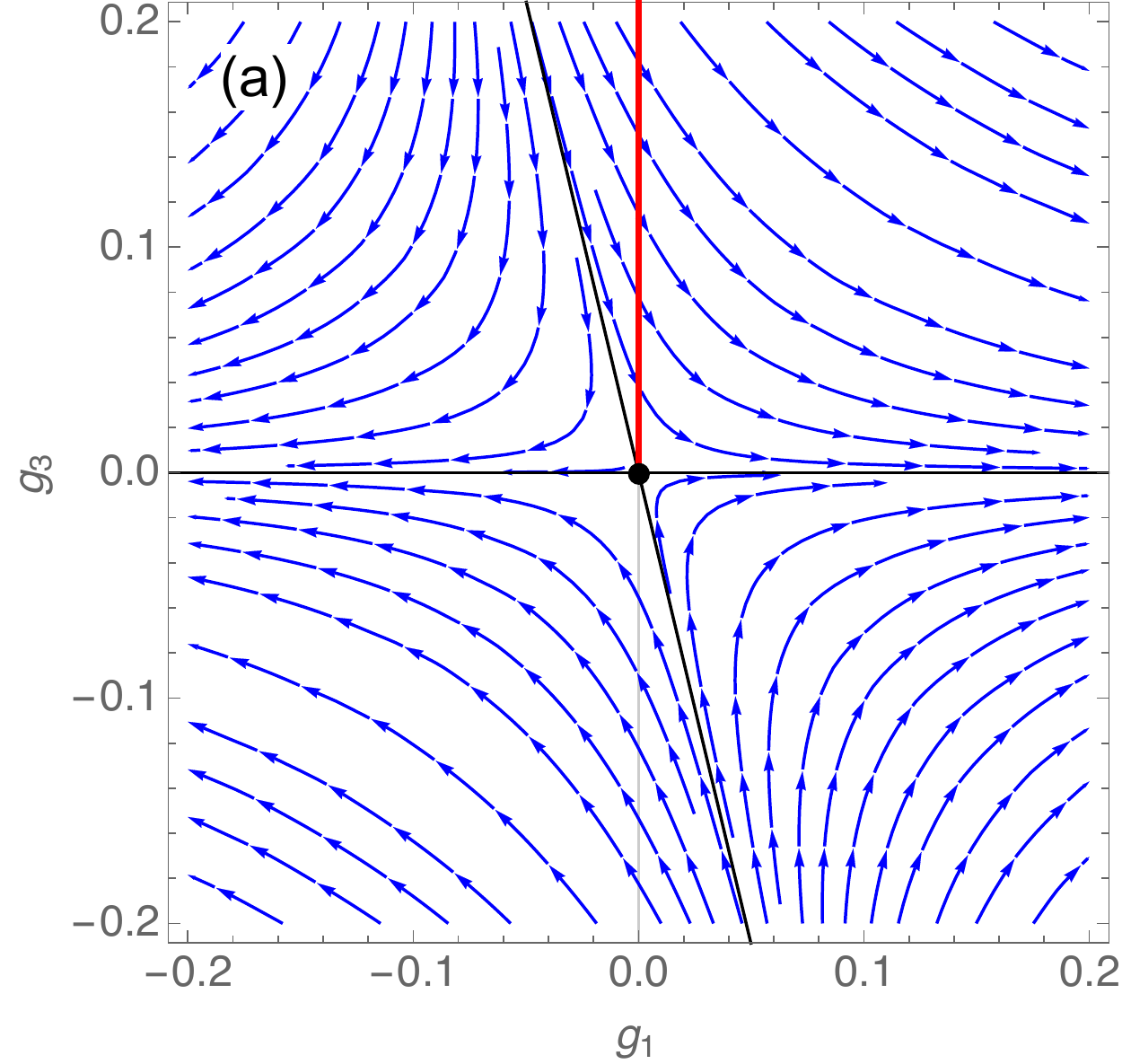}%
}\hfill
\subfloat{%
  \includegraphics[width=.4\linewidth]{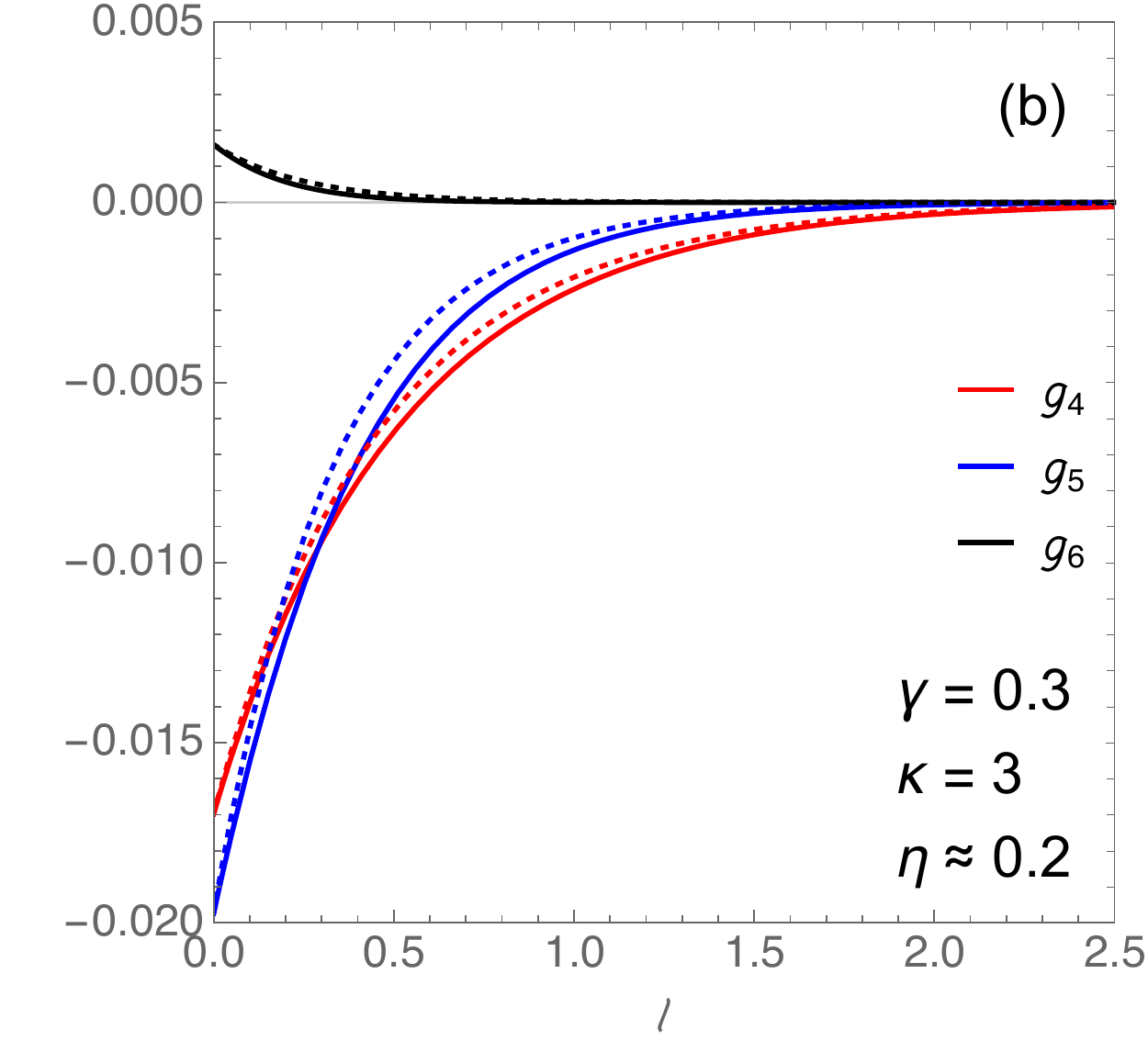}%
}
\caption{RG flow in the vicinity of the Gaussian fixed point. Panel (a): RG flow in the intersection of the subspace $\{g_1,g_3\}$ and the hyperplane $g_4 = g_5 = g_6 = 0$. Black dot marks the Gaussian fixed point. Thin black lines indicate the eigenvectors of the linearized flow and the red line shows possible initial conditions $g_{1}(0)= 0$, $g_{3}(0) > 0$ [see Eq. (\ref{RG_init_conds})]. Panel (b): RG flow of $g_4$, $g_5$, and $g_6$ as given by Eq. (\ref{RG_g}) [solid lines] and Eq. (\ref{RG_lin}) [dotted lines] for typical values of $\kappa$, $\gamma$, and $\eta$. The ratio $\Lambda/mv_F$ is $1$.}
\label{Gaussian_FP}
\end{figure}

\end{widetext}

\begin{widetext}

\begin{figure}[t]
\includegraphics[width=\linewidth]{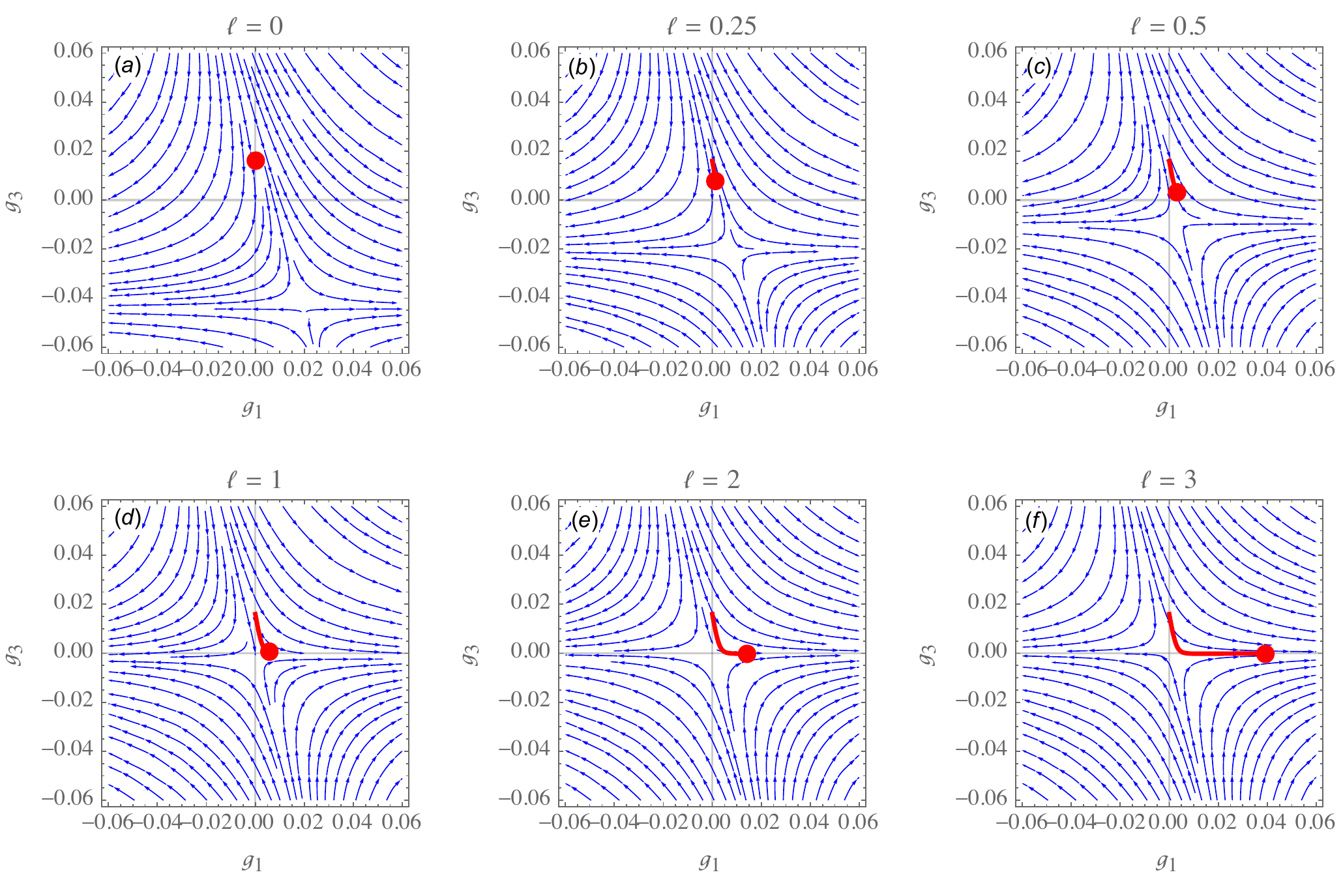}
\caption{A step-by-step RG flow in the intersection of the $\{g_1, g_3\}$ subspace and the hyperplane defined by the values of $g_4(l)$, $g_5(l)$, and $g_6(l)$ at a given $l$. The red dot represents the corresponding state of the system. Panel (a) shows the initial state of the system, $l = 0$ and all couplings $g_i$ having their initial values $g_{i,0}$, as given by Eq. (\ref{RG_init_conds}). Here we take typical values $\kappa = 3$, $\gamma = 0.3$, $\eta \approx 0.2$, and put $\Lambda/m v_F = 1$ [the same as in Fig.~\ref{Gaussian_FP} (b)]. In the consequent panels (b)--(f) the RG time $l$ increases, and the hyperplane $\{g_4(l), g_5(l), g_6(l)\}$ evolves according to the RG equations (\ref{RG_g}). This explains the variation of the flow lines from one panel to another [most noticeble in (a) -- (d)]. The red line shows the path travelled by the system as it flows from the initial state at $l = 0$ to the current state. After $l \gtrsim 1$, the couplings $g_4$, $g_5$, and $g_6$ are essentially zero, and in panels (d)--(f) the flow lines in the $\{ g_1, g_3 \}$ plane remain practically the same.}
\label{RG_flow_steps}
\end{figure}
	
\end{widetext}

Using the momentum-shell RG approach in $d=1$, we obtain the following one-loop flow equations (for the derivation see Appendix \ref{AppMomShellRG}):
\be \label{MomShellRG}
\begin{aligned}
	\partial_l A &= (2\chi -z + 1)A, \\
	\partial_l u_2 &= (2\chi + z-1) u_2 + 12 \, {\cal G}_1 \, u_4 - 18 \, {\cal G}_2 \, u_3^2,\\
	\partial_l u_3 &= (3\chi+z-2) u_3 + 10 \, {\cal G}_1 \, u_5 -  36 \, {\cal G}_2 u_3 u_4,\\
	\partial_l u_4 &= (4\chi+z-3)u_4 + 15 \, {\cal G}_1 \, u_6 - 36 \, {\cal G}_2 \, u_4^2 -  60 \, {\cal G}_2 \, u_3 u_5,\\
	\partial_l u_5 &= (5\chi+z-4) u_5 - 120 \, {\cal G}_2 \, u_4 u_5 - 90 \, {\cal G}_2 \, u_3 u_6, \\
	\partial_l u_6 &= (6\chi+z-5) u_6 - 100 \,  {\cal G}_2 \, u_5^2 - 180 \, {\cal G}_2 \, u_4 u_6,\\ 
\end{aligned}
\ee
where ${\cal G}_1 = (\Lambda^2/2\pi) (A\, u_2)^{-1/2}$ and ${\cal G}_2 = {\cal G}_1/(2u_2)$.
Importantly, RG procedure also generates a term $u_1 \Phi'_{\sigma}$, whose coupling flows according to
\be \label{RG_u_1}
	\partial_l u_1 = (\chi+z)u_1 + 3 \,{\cal G}_1 \, u_3.
\ee
Note that the above equation is decoupled from the rest of the RG equations (\ref{MomShellRG}).
For the $u_1$ term we define a corresponding dimensionless coupling as
\be \label{g_1}
	g_1 = \frac{A^{1/4} \, u_1}{\sqrt{\pi} \,\Lambda u_2^{3/4}}.
\ee
Unlike all other dimensionless couplings, $g_1$ is relevant. 
It is more convenient to study the RG flow in terms of the dimensionless couplings $g_n$. Using Eqs. (\ref{g_n}), (\ref{MomShellRG}), and (\ref{RG_u_1}), one can show that the couplings obey the equations
\be \label{RG_g}
\begin{aligned}
	\partial_l g_1 &= g_1+\frac{3}{2 \pi} g_3 - \frac{9}{2}  g_1 g_4 + \frac{27}{8} g_1 g_3^2,\\
	\partial_l g_3 &= - g_3 + \frac{5}{\pi} g_5 - \frac{33}{2} g_3 g_4 + \frac{45}{8} g_3^3,\\
	\partial_l g_4 &= - 2 g_4 + \frac{15}{2 \pi} g_6 - \frac{15}{\pi} g_3 g_5 -18 g_4^2 + \frac{27}{4} g_4 g_3^2,\\
	\partial_l g_5 &= - 3 g_5 -\frac{45}{2} g_6 g_3 - \frac{81}{2} g_4 g_5 + \frac{63}{8} g_5 g_3^2,\\
	\partial_l g_6 &= - 4 g_6 - \frac{25}{\pi}g_5^2 - 57 g_4 g_6 + 9 g_6 g_3^2,
\end{aligned}
\ee 
with $g_{1}(0) = 0$ and other initial conditions following from~Eq. (\ref{GL_coeffs}):
\be \label{RG_init_conds}
\begin{aligned}
	g_{3}(0)&= \frac{1 }{2 K_{\sigma}^{3/2} } \frac{\Lambda}{m v_F} \left(\frac{u_{\sigma}}{v_F}\right)^2 \frac{\eta^{5/2} \kappa^{3/2}}{ \left(2-y\right)^{5/4} \left(2+y/K_{\sigma}^2 \right)^{1/4} }, \\
	g_{4}(0)&= - \; \frac{ \Lambda^2}{8 m^2} \frac{K_{\rho}}{u_{\rho}} \frac{K_{\sigma}^2}{u_{\sigma}} \frac{(1-2y)}{ \left(2 - y \right)^{3/2} \left(2+y/K_{\sigma}^2 \right)^{1/2} },\\
	g_{5}(0)&= - \; \frac{3\pi}{8} \frac{K_{\rho} K_{\sigma}^{1/2} u_{\sigma}}{u_{\rho}} \frac{\Lambda^3}{\left(m v_F\right)^3} \frac{ \left(2 - y \right)^{-7/4} \eta^{5/2} \kappa^{3/2}}{  \left(2+y/K_{\sigma}^2 \right)^{3/4} } ,\\
	g_{6}(0)&= \frac{\pi}{32}\left(\frac{\Lambda^2 K_{\rho} K_{\sigma}^2}{m^2 u_{\rho} u_{\sigma}} \right)^2 \frac{(2-7y)}{\left(2 - y \right)^{2} \left(2 + y/K_{\sigma}^2 \right)}.
\end{aligned}
\ee
The harmonic coupling $u_2$ then flows according to
\be
	\partial_l u_2 = \left(2\chi + z - 1 + 6 \,g_4 - \frac{9}{2} \, g_3^2  \right) u_2.
\ee
We thus see that the Gaussian fixed point is unstable and for any nonzero values of $g_{3}(0)$ and $g_{5}(0)$ [i.e. for $\eta\neq0$, see Eq. (\ref{RG_init_conds})], the system flows away in the direction of $g_1$. This happens despite the initial condition for $g_1$ is strictly zero. A detailed analysis shows that for all realistic initial conditions, given by Eq. (\ref{RG_init_conds}), the flow is such that $g_1$ increases towards positive values, whereas all other couplings tend to zero. For a typical initial condition this is illustrated in Figs.~\ref{Gaussian_FP} and~\ref{RG_flow_steps}.  

Let us now analyze the system of RG equations in the vicinity of the Gaussian fixed point \cite{fixed_points}.
Eqs. (\ref{RG_g}) can be written in the matrix form as $\partial_l \boldsymbol{g} = \boldsymbol{\beta} (\boldsymbol{g})$.
The Jacobian matrix for the system linearized near the origin is
\be
	\boldsymbol{J} = \frac{\partial \left( \beta_1, \beta_3, \ldots, \beta_6 \right)}{\partial \left( g_1, g_3, \ldots, g_6 \right)}\Bigl|_0  = \begin{pmatrix}
			\,1 & \frac{3}{2\pi} &     &         &   \\
			 & -1         &     & \frac{5}{\pi} &   \\
			 &      	& -2 &        &\frac{15}{2\pi} \\
			 & 		&     & -3    &   \\
			 & 		&     &   	& -4 \,
		     \end{pmatrix}.		
\ee
Then, the solution of the linearized system is $\boldsymbol{g}(l) = e^{ l \boldsymbol{J} } \boldsymbol{g}_0$. In components it reads
\be \label{RG_lin}
\begin{aligned}
	g_1(l)&= \frac{3 }{2 \pi} \,g_{3}(0)\, \sinh{l} +\frac{15}{4 \pi ^2} \, g_{5}(0)\, e^{-l} \sinh^2{l},\\
	g_3(l)&= g_{3}(0) \, e^{-l}+\frac{5}{\pi}\, g_{5}(0)\, e^{-2 l} \sinh{l},\\
	g_4(l)&= g_{4}(0) \,e^{-2 l}+\frac{15}{2 \pi}\, g_{6}(0) \,e^{-3 l} \sinh{l},\\
	g_5(l)&= g_{5}(0) \,e^{-3 l},\\
	g_6(l)&= g_{6}(0)\,e^{-4 l}. 
\end{aligned}
\ee

Therefore, the system experiences a runaway flow, as seen in Fig. \ref{RG_flow_steps}. This situation is typical for the first order phase transition \cite{Fisher1982, Binder1987}. 

\section{Phase transition criterion} \label{S:PT}

In order to obtain the phase transition criterion, let us consider the renormalized action that reads:
\be \label{S_renorm}
\begin{aligned}
	S = \int dx d\tau &\left\{ \frac{A}{2} \dot\Phi_{\sigma}^2 + \frac{u_2}{2} \Phi_{\sigma}'{}^2 +u_4 \Phi_{\sigma}'{}^4 + u_6 \Phi_{\sigma}'{}^6  \right. \\
	&+ u_1 \Phi_{\sigma}' + u_3 \Phi_{\sigma}'{}^3 + u_5 \Phi_{\sigma}'{}^5 \Bigr\},
\end{aligned}
\ee
where the coefficients $u_j = u_j(l)$ are the solutions to RG equations (\ref{MomShellRG}), and are related to the dimensionless couplings $g_j$ via Eqs. (\ref{g_n}) and (\ref{g_1}).

Before we proceed, let us make an important remark. One should keep in mind that in the context of ultracold atomic gases, magnetization is nothing else than the difference between the number of atoms in the (pseudo)spin-$\uparrow$ and -$\downarrow$ states. This means that the population of each spin species is conserved. Restricting ourselves to the case with no population imbalance, the {\it total} magnetization is zero in all phases:
\be \label{tot_magn_zero}
	\int dx\, \Phi_{\sigma}'(x) = 0.
\ee
Therefore, the linear term in action (\ref{S_renorm}) should be discarded. Note that although condition (\ref{tot_magn_zero}) prohibits phases with nonzero total magnetization, it still allows the existence of spin configurations, where the magnetization is different from zero {\it locally}. In other words, one can have a system of domains.

At sufficiently large RG times $l$, in order to find the phase transition, one may simply minimize the renormalized Hamiltonian density \cite{Kozii2017}. The latter follows from Eq. (\ref{S_renorm}) and reads
\be \label{coarse_grained_H}
	{\cal H} = \frac{u_2}{2} \Phi_{\sigma}'{}^2 + u_3 \Phi_{\sigma}'{}^3 -|u_4| \Phi_{\sigma}'{}^4 + u_5 \Phi_{\sigma}'{}^5 + u_6 \Phi_{\sigma}'{}^6,
\ee
where we omitted the kinetic term $\Pi_{\sigma}/2A$.
At $l=0$ coefficients $u_3$ and $u_5$ are $\sim (\eta \kappa)^{5/2}$, and they are negligibly small for $\eta \kappa \ll 1$ [see Eq. (\ref{GL_coeffs})]. At larger $l$ these coefficients become even smaller. For this reason, we can neglect cubic and quintic terms in the Hamiltonian (\ref{coarse_grained_H}). 
Thus, up to small corrections, the phase transition criterion is
\be \label{PT_crit}
	\frac{u_2 (l) u_6(l)}{u_4^2(l)} \approx \frac{1}{2}.
\ee
In the region, where $u_2 u_6 / u_4^2 < 1/2$, the system is in the phase with zero magnetization $\langle \Phi' \rangle$, whereas for $u_2 u_6 / u_4^2 > 1/2$ the ground state has $\langle \Phi' \rangle \neq 0$.

Using Eq. (\ref{g_n}) we express $u_4$ and $u_6$ in terms of the dimensionless couplings $g_4$ and $g_6$, and equation (\ref{PT_crit}) becomes $g_6(l)/g_4^2(l) = \pi/2$. Here $g_4(l)$ and $g_6(l)$ are the solutions of RG equations (\ref{RG_g}). In the viscinity of the Gaussian fixed point the RG flow is such that it is well described by the linearized RG equations with the solutions (\ref{RG_lin}), as can be seen from Fig. \ref{Gaussian_FP}(b). Thus, the phase transition criterion (\ref{PT_crit}) can be written as
\be \label{PT_crit_g}
	 \frac{g_{6}(0)}{g_{4}^2(0)} \; \frac{1}{ 1 - \delta(l) } = \frac{\pi}{2},
\ee
where the bare couplings $g_{4}(0)$ and $g_{6}(0)$ are given by Eq.~(\ref{RG_init_conds}), and we introduced the quantity
\be \label{delta_def}
	\delta(l) = \frac{\pi \Lambda^2}{4 (m v_F)^2} \frac{(2-7y) \, e^{-l} \sinh l}{\left(2 - y \right)^{1/2} \left(2 + y/K_{\sigma}^2 \right)^{1/2} (1-2y)}.
\ee
From Eqs. (\ref{PT_crit_g}) and (\ref{delta_def}) we see that the distance to the phase transition is controlled by the parameter $y$, given by Eq. (\ref{y_def}).

At $l =0$ the correction $\delta(l)$ is zero and Eq. (\ref{PT_crit_g}) reduces to
\be \label{MF_transition}
	\frac{(1-y/2) (1-7y/2)}{(1-2y)^2} = \frac{1}{16},
\ee
where we used Eq. (\ref{RG_init_conds}) for $g_{4}(0)$ and $g_{6}(0)$. Eq. (\ref{MF_transition}) is essentially the mean field phase transition criterion, since it involves only the bare couplings. The roots of Eq. (\ref{MF_transition}) are $y = (5 \pm \sqrt{15})/4$. Taking the minus sign we have $y_{*} \approx 0.282$. Thus, at the mean field level there is a first order phase transition at $y_* = 0.282$, which can be the case at fairly weak coupling.

At finite values of $l$ and for $y$ sufficiently close to $y_*$, one has $\delta(l) \ll 1$. This is because $\Lambda / m v_F \sim 1$ and the $l$-dependent factor in Eq. (\ref{delta_def}) is bounded by $1/2$.  We then look for the solution of Eq. (\ref{PT_crit_g}) in the form $y = y_* + \delta y$ and obtain
\be
	\delta y \sim 10^{-4} ( \Lambda / m v_F)^2 e^{-l} \sinh l \sim 10^{-4}.
\ee
 We see that the correction $\delta y$ to the mean field critical value $y_* \approx 0.282$ is negligibly small. 
One may easily check that in this case the correction $\delta(l)$ is also negligible, being of the order of $10^{-2}$.

Thus, the phase transition criterion is given by Eq.~(\ref{MF_transition}), and the RG corrections can be neglected.
The spontaneous magnetization in this case is
\be \label{magn_MF}
	{\cal M} \equiv \langle \Phi_{\sigma}' \rangle = \pm \sqrt{u_2 / |u_4|}.
\ee
Importantly, there are two minima, and hence it is possible to have an instanton-like field configuration that tunnels from one minimum to another, creating a sequence of domains (see, e.g., \cite{Rajaraman}).

\begin{figure}[t]
\includegraphics[width=\linewidth]{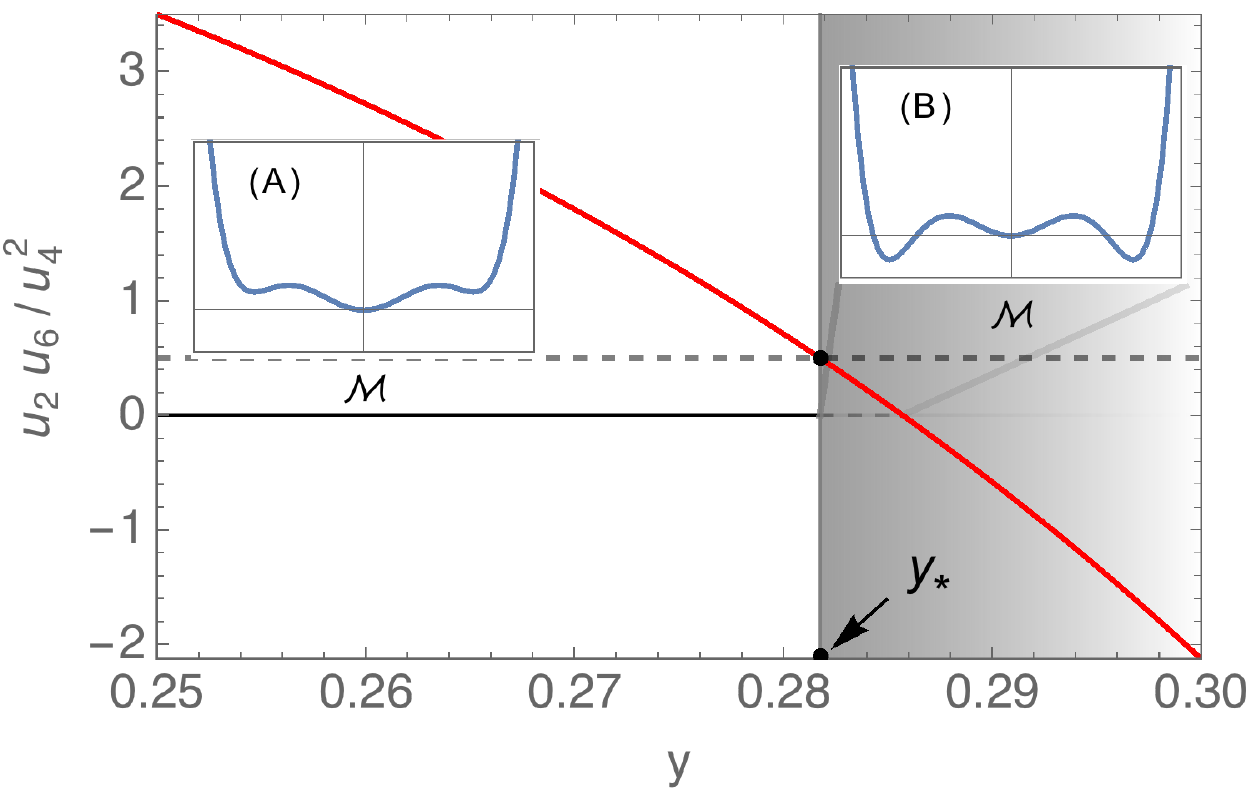}%
\caption{Phase diagram as follows from Eq. (\ref{MF_transition}). The red line shows the ratio $g_{6}(0) /(\pi  g_{4}(0)^2)$ versus the dimensionless parameter $y$, given by Eq. (\ref{y_def}). There is a critical value $y_* \approx 0.282$, where $g_{6}(0) /(\pi  g_{4}(0)^2) = 1/2$ and  the phase transition occurs. For $y<y_*$, the system is in the paramagnetic phase and the magnetization ${\cal M}$ is zero, as illustrated schematically on the inset (A). For $y>y_*$, the system enters the phase with the nonzero (local) magnetization ${\cal M}$, as shown on the inset~(B).}
\label{F:MF_transition}
\end{figure}

As illustrated in Fig. \ref{F:MF_transition}, for $y< y_* \approx 0.282$ the system is in the phase with ${\cal M} = 0$, whereas for $y > y_*$ the phase with a nonzero ${\cal M}$ has a lower energy. 

\section{Discussion and Conclusions} \label{S:concl}

To summarize, in this paper we considered a zero temperature one-dimensional two-component Fermi gas with a weak/intermediate contact repulsive interaction in the even-wave channel and an additional attractive odd-wave interaction between particles in the spin-$\uparrow$ state. Using bosonization technique we derived an effective field theory for the spin degrees of freedom, described by the Lagrangian (\ref{Lagr}). 

In the regime of weak/intermediate even-wave repulsion ($\gamma \lesssim 1$) and a near-resonant odd-wave attraction, we have found a first order phase transition to a state with a nonzero local magnetization. 
The distance to the phase transition is controlled by the dimensionless parameter $y$, given by Eq. (\ref{y_def}). The phase transition occurs at the critical value $y_* \approx 0.282$. At smaller values the system is in the phase with zero magnetization. At larger values, in the region $y > y_*$, the system enters the phase with a nonzero local magnetization.

The phase with nonzero local magnetization deserves comments. In the context of ultracold atoms, the spin-$\uparrow$ and spin-$\downarrow$ states of a fermion are, actually, two distinct atomic hyperfine states. Magnetization in this language is then simply the difference between the populations of atoms in these states. In the absence of inelastic collisions this difference remains constant.
Thus, when going through the phase transition, magnetization can change only locally, whereas the total magnetization remains zero. This is nothing else than the development of domains. In each domain there are more atoms in one hyperfine state than in the other. Therefore, locally, the magnetization is different from zero.

\begin{acknowledgements}
We thank E. Demler, F. Essler, O. Gamayun, D.~Petrov, and M.~Zvonarev for useful discussions, and M.~Zaccanti for drawing our attention to Refs. \cite{Valtolina2017, Scazza2017, Amico2018}.
The research leading to these results has received funding from the European Research Council under European Community's Seventh Framework Programme (FR7/2007-2013 Grant Agreement no. 341197).
This work is part of the Delta-ITP consortium, a program of the Netherlands Organization for Scientific Research (NWO) that is funded by the Dutch Ministry of Education, Culture and Science (OCW).
\end{acknowledgements}

\appendix 
\begin{widetext}
\section{The integral in Eq. (\ref{Isp})} \label{A:integrals}
In this Appendix we calculate the integral of the form
\be \label{I_sp_int}
	\tilde {\cal I}_{s,p} =  \int_0^{+\infty} dx\, \int_0^{+\infty} dy \,e^{-\frac{x^2}{y} - \tilde \nu y }  \frac{ x^s y^{p-\frac{1}{2}} }{\left( x^2 + y^2 +a^2 \right)^K} \qquad (K>0),
\ee
which is a slight generalization of the integral in Eq. (\ref{Isp}). Using the well known identity 
\be
	\frac{1}{\left( x^2 + y^2 +a^2 \right)^K} = \frac{1}{\Gamma(K)} \int_0^{+\infty} dq \, q^{K-1} e^{-\left( x^2 + y^2 +a^2 \right)\, q},
\ee
we write
\be
	\tilde {\cal I}_{s,p} =
	\frac{1}{\Gamma\left( K \right)} 
	\int_0^{+\infty} dq \, e^{-a^2 q} q^{K-1}
	\int_0^{+\infty} dy \,y^{p-\frac{1}{2}} \, e^{-\tilde \nu y - q y^2}   
	\int_0^{+\infty} dx \, x^s \,e^{- \left( \frac{1}{y} +q \right) x^2 } .
\ee
The integral over $dx$ gives
\be
	\frac{1}{2}\Gamma\left(\frac{s+1}{2}\right) \left( q + \frac{1}{y} \right)^{-(s+1)/2}
\ee
Then, taking the limit $a \to 0$ and making a change of variables $z = q y$ we obtain
\be \label{I_sp_int}
	\tilde {\cal I}_{s,p} =  
	\frac{1}{2} \Gamma\left(\frac{s+1}{2}\right) \int_0^{+\infty} dy \, e^{-\tilde \nu y} \,y^{\frac{s}{2} + p - K}
	\frac{1}{\Gamma\left( K \right)}\int_0^{+\infty} dz \, e^{-y z} \, z^{K-1}  \left( 1 + z \right)^{-(s+1)/2}.
\ee
In the integral over $dz$ one recognizes the integral representation of the Tricomi hypergeometric function. Its general form reads
\be
	U(\alpha,\beta,y) = \frac{1}{\Gamma(\alpha)} \int_0^{+\infty} dz\, e^{-y z} z^{\alpha-1} (1+z)^{\beta-\alpha-1},
\ee
which is valid for $\text{Re}\, y >0$ and $\text{Re}\, \alpha > 0$. One has the following expression for $U(\alpha,\beta,y)$ in terms of the confluent hypergeometric function $_1 F_1(\alpha,\beta,y)$:
\be
	U(\alpha,\beta,y) = \frac{\Gamma(1-\beta)}{\Gamma(\alpha-\beta+1)} \, _1 F_1(\alpha,\beta,y)  
	+ \frac{\Gamma(\beta-1)}{\Gamma(\alpha)} \, y^{1-\beta} \, _1 F_1(\alpha-\beta+1,2-\beta,y).
\ee
In our case we have $\alpha = K$ and $\beta = 1 + K - (s+1)/2$, which yields
\be
\begin{aligned}
	&\frac{1}{\Gamma\left( K \right)}\int_0^{+\infty} dz \, e^{-y z} \, z^{K-1}  \left( 1 + z \right)^{-\frac{s+1}{2}} = U(K, 1 + K - \frac{s+1}{2}, y) \\
	&= \frac{\Gamma\left( \frac{s+1}{2} - K \right)}{\Gamma\left( \frac{s+1}{2} \right)} \,_1 F_1 \left( K, 1+ K - \frac{s+1}{2}, y \right) + \frac{\Gamma\left( K-\frac{s+1}{2} \right)}{\Gamma\left( K \right)} \, y^{\frac{s+1}{2}-K} \,_1 F_1 \left( \frac{s+1}{2}, 1+ \frac{s+1}{2} -K, y \right)
\end{aligned}
\ee
Then, $\tilde{\cal I}_{s,p}$ becomes
\be
	\tilde {\cal I}_{s,p} = \frac{1}{2}\Gamma\left( \frac{s+1}{2} - K \right) J_{s,p}^{(1)} + \frac{\Gamma\left( \frac{s+1}{2}\right) \Gamma\left( K- \frac{s+1}{2} \right)}{2\Gamma\left( K \right)} J_{s,p}^{(2)},
\ee
where we defined
\be
\begin{aligned}
	J_{s,p}^{(1)} &= \int_0^{+\infty} dy\, e^{-\tilde\nu y} y^{\frac{s}{2}+p - K} \, _1 F_1\left(K,\, 1+K-\frac{s+1}{2},\, y\right),\\
	J_{s,p}^{(2)} &= \int_0^{+\infty} dy\, e^{-\tilde\nu y} y^{s+p+\frac{1}{2}- 2K }  \, _1 F_1\left(\frac{s+1}{2},\, 1- K+\frac{s+1}{2},\, y\right).
\end{aligned}
\ee
The above integrals can be calculated in terms of the ordinary hypergeometric function $_2 F_1$ using \cite{LLvol3}
\be
	\int_0^{+\infty} dy \, e^{-\lambda y} y^{\nu} \, _1F_1(\alpha,\beta,k y) = \Gamma(\nu+1) \lambda^{-\nu-1} \, _2 F_1(\alpha, \nu+1, \beta, k/\lambda),
\ee
provided that $\text{Re} \, \nu > -1$ and $\text{Re} \, \lambda > |\text{Re}\, k|$. In the case $0<K \leq 1$ both these conditions are satisfied for $J_{s,p}^{(1)}$ and $J_{s,p}^{(2)}$ with any relevant combination of non-negative integers $s$ and $p$. Therefore, we finally obtain
\be
\begin{aligned}
	J_{s,p}^{(1)} &=\Gamma\left(\frac{s}{2}+p+1- K \right) \tilde{\nu}^{-\left(p+\frac{s}{2} - K+1\right)} \, _2F_1 \left( K, \, \frac{s}{2} + p - K + 1, 1 + K - \frac{s+1}{2}, \frac{1}{\tilde{\nu}} \right), \\
	J_{s,p}^{(2)} &=\Gamma\left(s+p+\frac{3}{2} - 2 K \right) \tilde{\nu}^{-\left(p+s+\frac{3}{2} - 2 K \right)} \, _2F_1 \left( \frac{s+1}{2}, \, s + p + \frac{3}{2} - 2 K, 1 - K  + \frac{s+1}{2}, \frac{1}{\tilde{\nu}} \right).
\end{aligned}
\ee
Putting $K = 1$ and $\tilde \nu = 1/(\eta \kappa)$, we get
\be
\begin{aligned}
	\tilde{\cal I}_{s,p} &= \frac{1}{2} \Gamma \left(\frac{s+1}{2}-1\right) (\eta \kappa)^{p+\frac{s}{2}} \Gamma
   \left(p+\frac{s}{2}\right) \, _2F_1\left(1,p+\frac{s}{2};\frac{1}{2} (-s-1)+2;\eta
   \kappa \right)\\
   &+\frac{1}{2} \Gamma \left(\frac{1-s}{2}\right) \Gamma
   \left(\frac{s+1}{2}\right) \Gamma \left(p+\frac{s-1}{2}\right) (\eta \kappa)^{p+s-\frac{1}{2}} (1-\eta \kappa
   )^{-p-\frac{s}{2}+\frac{1}{2}}.
\end{aligned}
\ee
For $\eta\kappa \ll 1$ it behaves as
\be
	\tilde{\cal I}_{s,p} \approx 
		\begin{cases}
			\frac{1}{2} \Gamma\left( \frac{s}{2} + p \right) \Gamma\left( \frac{s - 1}{2} \right) (\eta\kappa)^{\frac{s}{2}+p}, \quad s\neq 0; \\
			\frac{\pi}{2} \Gamma\left( p - \frac{1}{2} \right) (\eta\kappa)^{p - \frac{1}{2}}, \;\,\quad\quad\qquad s=0,
		\end{cases}
\ee
which leads to Eqs. (\ref{alpha_s_p}) and (\ref{alpha_s_p2}) in the main text.

\section{Bosonized odd-wave interaction} \label{A:bosonization}
In this Appendix we present the terms that we keep after expanding the $\Theta_{\rho, \sigma}$ fields and the normal-ordered exponent in Eq.~(\ref{norm_ord_exp_spin_charge}).
In the charge sector these are the terms up to the second order in the fields:
\be \label{Lcharge}
	\delta {\cal L}_{\text{odd}}^{(\rho)} = - i \sqrt{2\pi} k_F \alpha_{0,1} \dot \Theta_{\rho} + \pi v_F^{-1}\alpha_{0,2} \dot \Theta_{\rho}^2 + \pi v_F \alpha_{2,0} \Theta_{\rho} '{}^2.
\ee
In the spin sector -- up to the sixth order:
\be \label{Lspin}
\begin{aligned}
		&\delta {\cal L}_{\text{odd}}^{(\sigma)} = - i \sqrt{2\pi} k_F \alpha_{0,1} \dot \Theta_{\sigma} 
		+ \pi v_F^{-1}\alpha_{0,2} \dot \Theta_{\sigma}^2 + \pi v_F \alpha_{2,0} \Theta_{\sigma} '{}^2
		+ \frac{\sqrt{2} \pi ^{3/2} }{3 m v_F^3} \alpha _{0,3}\, i\, \dot\Theta_{\sigma}^3 
		-\frac{\pi}{4 m^2 v_F^3} \alpha _{2,2} \dot\Theta_{\sigma}'{}^2\\
		&  
		-\frac{\pi ^2 }{6 m^2 v_F^5} \alpha _{0,4} \dot\Theta_{\sigma}^4 -\frac{\pi ^2}{6 m^2 v_F}  \alpha _{4,0} \Theta_{\sigma}'{}^4
		-\frac{\pi ^{5/2}}{15 \sqrt{2} m^3 v_F^7}  \alpha _{0,5} \,i\,\dot\Theta_{\sigma}^5
		+ \frac{\pi ^3}{90 m^4 v_F^9}  \alpha _{0,6}\dot \Theta_{\sigma}^6
		+\frac{\pi ^3}{90 m^4 v_F^3}\alpha _{6,0} \Theta_{\sigma}'{}^6.
\end{aligned}
\ee
Finally, we also keep the following terms that couple spin and charge:
\be \label{Lspincharge}
	\delta{\cal L}_{\text{odd}}^{(\rho\sigma)} = 
	2\pi v_F \alpha_{2,0} \Theta'_{\rho} \Theta'_{\sigma} 
	+ \frac{2\pi }{v_F} \alpha_{0,2} \dot\Theta_{\rho} \dot\Theta_{\sigma} 
	+ \frac{\sqrt{2} \pi ^{3/2} }{m v_F^3} \alpha _{0,3} i \dot\Theta_{\rho} \dot\Theta_{\sigma}^2
	+\frac{\sqrt{2} \pi ^{3/2}}{m v_F} \alpha _{2,1} i \dot\Theta_{\rho} \Theta_{\sigma}'{}^2.
\ee

\section{Effective Lagrangian for the spin fields}
\label{spin_fields_eff_action}

In this Appendix we integrate out the charge fields and obtain an effective Lagrangian for the spin degrees of freedom, given by Eq. (\ref{Lagr}) in the main text.

\subsection{Total bosonized Lagrangian}

Combining Eqs. (\ref{Lcharge}), (\ref{Lspin}), and (\ref{Lspincharge}) with Eq. (\ref{L_YG}), we arrive to the total Lagrangian ${\cal L} = {\cal L}_{\sigma} + {\cal L}_{\rho}^{(0)} + {\cal L}_{\rho\sigma}$ given by Eqs. (\ref{L_sigma}) -- (\ref{L_rho_sigma}) in the main text. The coefficients in the Lagrangian are given by
\be \label{coeffs}
\begin{aligned}
	&\varepsilon_{\rho} = \varepsilon_{\sigma} = \sqrt{2\pi} k_F \alpha_{0,1} \approx (\sqrt{2}/\pi) k_F \sqrt{\eta/\kappa} \\
	&\quad A_2^{\rho,\sigma} = \frac{K_{\rho,\sigma}}{u_{\rho,\sigma}} + \frac{2\pi \alpha_{0,2}}{v_F}, \quad
	A_3^{\sigma} = \frac{\sqrt{2} \pi ^{3/2} \alpha _{0,3}}{3 m v_F^3}, \quad
	A_4^{\sigma} = \frac{\Gamma ^2}{8} \frac{K_{\rho}}{u_{\rho}} \frac{K_{\sigma}^4}{u_{\sigma }^4} + \frac{\pi^2 \alpha _{0,4} }{6 m^2 v_F^5}, \quad
	A_5^{\sigma} = \frac{\pi ^{5/2} \alpha _{0,5}}{15 \sqrt{2} m^3 v_F^7},\quad
	A_6^{\sigma} = \frac{\pi^3 \alpha _{0,6}}{90 m^4 v_F^9},\quad \\
	&B_2^{\rho,\sigma} = u_{\rho,\sigma} K_{\rho,\sigma} + 2\pi v_F \alpha_{2,0},\quad 
	B_4^{\sigma} = \frac{\Gamma ^2}{8} \frac{K_{\rho}}{u_{\rho}} +\frac{\pi^2 \alpha _{4,0}}{6 m^2 v_F}, \quad 
	B_6^{\sigma} = \frac{\pi^3 \alpha _{6,0}}{90 m^4 v_F^3},\\
	& C = 2\pi v_F \alpha_{2,0} , \quad 
	D = \frac{2\pi}{v_F} \alpha_{0,2}, \quad
	E = \frac{\Gamma}{2} \frac{K_{\rho}}{u_{\rho} } + \frac{\sqrt{2} \pi ^{3/2} \alpha _{2,1}}{m v_F}, \quad 
	F = \frac{\Gamma}{2} \frac{K_{\rho}}{u_{\rho} } \frac{K_{\sigma}^2}{u_{\sigma}^2} - \frac{\sqrt{2} \pi ^{3/2} \alpha _{0,3}}{m v_F^3}.
\end{aligned}
\ee
After the shift $\Theta_{\rho} = \tilde\Theta_\rho + i \beta_{\rho} \tau$, given by Eq. (\ref{shift}), the Lagrangian becomes
\be \label{L_sigma_shift}
	{\cal L}_{\sigma} = i a_1^{\sigma} \dot{\Theta}_{\sigma} 
	+ \frac{a_2^{\sigma}}{2} \dot{ \Theta}_{\sigma}^2 
	+ \frac{b_2^{\sigma}}{2}  \Theta_{\sigma}'^2 
	+ a_3^{\sigma}\, i\, \dot{\Theta}_{\sigma}^3
	-a_4^{\sigma} \dot{\Theta}_{\sigma}{}^4 
	- b_4^{\sigma} \Theta_{\sigma}'{}^4 
	-a_5^{\sigma} \,i\,\dot{\Theta}_{\sigma}^5
	+ a_6^{\sigma} \dot{ \Theta}_{\sigma}^6
	+b_6^{\sigma} \Theta_{\sigma}'{}^6, 
\ee
\be \label{L_rho_shift}
	{\cal L}^{(0)}_{\rho} =  \frac{A_2^{\rho}}{2} \dot{\tilde \Theta}_{\rho}^2 + \frac{B_2^{\rho}}{2} \tilde \Theta_{\rho}'^2,
\ee
\be \label{L_rhosigma_shift}
	{\cal L}_{\rho\sigma} = C \tilde\Theta_{\rho}' \Theta_{\sigma}' + D \dot{\tilde \Theta}_{\rho} \dot{ \Theta}_{\sigma} + i \, \dot{\tilde{\Theta}}_{\rho} \Bigl\{ E {\Theta }_{\sigma}'{}^2 - F\dot{{\Theta }}_{\sigma }^2 \Bigr\},
\ee
where
\be \label{coeffs_shift}
\begin{aligned}
	& a_1^{\sigma} = D \beta_{\rho} - \varepsilon_{\sigma} = -\frac{K_{\rho}}{u_{\rho}} \beta_{\rho},\quad
	a_2^{\sigma} = A_2^{\sigma} + 2 F \beta _{\rho }, \qquad a_3^{\sigma} = A_3^{\sigma}, \quad
	a_4^{\sigma} = A_4^{\sigma}, \qquad 
   	a_5^{\sigma} =  A_5^{\sigma },\qquad
	a_6^{\sigma} = A_6^{\sigma} \\
	&b_2^{\sigma} =  B_2^{\sigma} -2 E \beta _{\rho}, \quad b_4^{\sigma} = B_4^{\sigma}, \quad b_6^{\sigma} = B_6^{\sigma}.
\end{aligned}
\ee
Since primarily we are interested in the spin sector, we proceed with integrating out the charge fields.

\subsection{Integration over the charge degrees of freedom}

The partition function can be written as
\be
	{\cal Z} = \int {\cal D} \Theta_{\sigma}\,e^{-\int dx d\tau {\cal L}_{\sigma} } \int{\cal D}\tilde \Theta_{\rho} \,e^{-\int dx d\tau {\cal L}_{\rho}^{(0)} -\int dx d\tau {\cal L}_{\rho \sigma}},
\ee
where the Lagrangian is given by Eqs. (\ref{L_sigma_shift}) -- (\ref{L_rhosigma_shift}).
The integral over ${\cal D}\tilde\Theta_{\rho}$ is Gaussian and, formally, can be calculated exactly. We first write the action corresponding to ${\cal L}_{\rho}^{0}$ as
\be
	S_{\rho}^{(0)} = \frac{1}{2}\int dx d\tau\, \tilde\Theta_{\rho}(x,\tau) \left\{- B_2^{\rho}\left( \partial_x^2+\frac{A_2^{\rho}}{B_2^{\rho}} \partial^2_{\tau} \right) \right\} \tilde\Theta_{\rho}(x,\tau) = \frac{1}{2}\int d1 d2 \tilde\Theta_{\rho}(1) G^{-1}_{\rho}(1-2)\tilde \Theta_{\rho}(2),
\ee
where $d1 = dx_1d\tau_1$, $d2 = dx_2d\tau_2$, and the Green's function is
\be
	G_{\rho}(x,\tau) = -\frac{1}{4\pi\sqrt{A_2^{\rho}B_2^{\rho}}} \ln \frac{x^2 + (B_2^{\rho}/A_2^{\rho}) \tau^2 + a^2}{a^2}.
\ee
The action corresponding to ${\cal L}_{\rho\sigma}$, i.e. terms that couple spin and charge, we write as
\be
	S_{\rho\sigma} = \int dx d\tau\tilde \Theta_{\rho} \left[ - C \Theta_{\sigma}'' - D \ddot{\Theta}_{\sigma} - i\,\partial_{\tau}\left( E \Theta_{\sigma}'^2 - F \dot{ \Theta}_{\sigma}^2 \right) \right] \equiv \int dx d\tau \tilde\Theta_{\rho} {\cal J}_{\sigma}.
\ee
Then, the integral over ${\cal D}\tilde\Theta_{\rho}$ takes the standard form and yields
\be
	\int {\cal D}\tilde\Theta_{\rho} \exp\left\{-\frac{1}{2} \int d1 d2 \tilde\Theta_{\rho}(1) G^{-1}_{\rho}(1-2) \tilde\Theta_{\rho}(2) + \int d1 \tilde\Theta_{\rho}(1) {\cal J}_{\sigma}(1)\right\} \propto \exp\left\{\frac{1}{2} \int d1 d2 {\cal J}_{\sigma}(1) G_{\rho}(1-2) {\cal J}_{\sigma}(2) \right\}.
\ee
Thus, integration over the charge degrees of freedom provides the following correction to $S_{\sigma}$:
\be
	\delta S = - \frac{1}{2} \int d1 d2 {\cal J}_{\sigma}(1) G_{\rho}(1-2) {\cal J}_{\sigma}(2) \to \sum_{j=1}^{4}\delta S_j,
\ee
where we kept only the most relevant terms:
\be
\begin{aligned}
	\delta S_1 &= -\frac{C^2}{2} \int d1 d2 \, \partial_{x_1}^2 \Theta_{\sigma}(1) G_{\rho}(1-2)\,\partial_{x_2}^2 \Theta_{\sigma}(2) = +\frac{C^2}{2} \int d1 d2 \, \partial_{x_1} \Theta_{\sigma}(1) G''_{\rho}(1-2)\,\partial_{x_2} \Theta_{\sigma}(2),\\
	\delta S_2 &= -\frac{D^2}{2} \int d1 d2 \, \partial_{\tau_1}^2 \Theta_{\sigma}(1) G_{\rho}(1-2)\,\partial_{\tau_2}^2 \Theta_{\sigma}(2) = +\frac{D^2}{2} \int d1 d2 \, \partial_{\tau_1} \Theta_{\sigma}(1) \ddot G_{\rho}(1-2)\,\partial_{\tau_2} \Theta_{\sigma}(2), \\
	\delta S_3 &= +\frac{E^2}{2} \int d1 d2 \, \partial_{\tau_1}\left(\partial_{x_1}\Theta_{\sigma}(1)\right)^2 G_{\rho}(1-2) \, \partial_{\tau_2}\left(\partial_{x_2}\Theta_{\sigma}(2)\right)^2 \\
	&= -\frac{E^2}{2} \int d1 d2 \, \left(\partial_{x_1}\Theta_{\sigma}(1)\right)^2 \ddot G_{\rho}(1-2) \, \left(\partial_{x_2}\Theta_{\sigma}(2)\right)^2,\\
	\delta S_4 &= +\frac{F^2}{2} \int d1 d2 \, \partial_{\tau_1}\left(\partial_{\tau_1}\Theta_{\sigma}(1)\right)^2 G_{\rho}(1-2) \, \partial_{\tau_2}\left(\partial_{\tau_2}\Theta_{\sigma}(2)\right)^2 \\
	&= -\frac{F^2}{2}\int d1 d2 \, \left(\partial_{\tau_1}\Theta_{\sigma}(1)\right)^2 \ddot G_{\rho}(1-2) \, \left(\partial_{\tau_2}\Theta_{\sigma}(2)\right)^2.
\end{aligned}
\ee
Using the results of subsection \ref{A:Integrate_out_charges}, for the above terms we obtain:
\be \label{deltaS_j}
\begin{aligned}
	\delta S_1 &= - \frac{C^2}{4B_2^{\rho}} \int d\tau dx \, \Theta_{\sigma}'^2, \qquad\qquad\qquad \delta S_2 = - \frac{D^2}{4A_2^{\rho}} \int d\tau dx \, \dot{\Theta}_{\sigma}^2, \\
	 \delta S_3 &= + \frac{E^2}{4A_2^{\rho}} \int d\tau dx \, \Theta_{\sigma}'^4, \qquad\qquad \delta S_4 = + \frac{F^2}{4A_2^{\rho}} \int d\tau dx \, \dot{\Theta}_{\sigma}^4.
\end{aligned}
\ee

Therefore, combining the above corrections with Eq. (\ref{L_sigma_shift}), the dual filed representation of the effective Lagrangian for the spin degrees of freedom becomes:
\be \label{L_sigma_eff}
	\tilde{\cal L}_{\sigma}= i a_1 \dot{\Theta}_{\sigma} +  \frac{a_2}{2} \dot{\Theta}_{\sigma}^2 + \frac{b_2}{2} \Theta_{\sigma}'^2 + i \frac{a_3}{3} \dot{\Theta}_{\sigma}^3 - \frac{a_4}{4} \dot{\Theta}_{\sigma}^4  - \frac{b_4}{4} \Theta_{\sigma}'^4 - i \frac{a_5}{5} \dot{\Theta}_{\sigma}^5 + \frac{a_6}{6} \dot{\Theta}_{\sigma}^6 + \frac{b_6}{6} \Theta_{\sigma}'{}^6,
\ee
where we defined
\be
\begin{aligned}
	&a_1 = a_1^{\sigma}, \quad a_2 = a_2^{\sigma} - \frac{D^2}{2A_2^{\rho}}, \quad b_2 = b_2^{\sigma} - \frac{C^2}{2B_2^{\rho}},\quad a_3 = 3 a_3^{\sigma}, \quad  a_4 = 4 a_4^{\sigma} - \frac{F^2}{A_2^{\rho}}, \quad  b_4 = 4 b_4^{\sigma} - \frac{E^2}{A_2^{\rho}}, \quad a_5 = 5a_5^{\sigma},\\
	&a_6 = 6 a_6^{\sigma}, \quad b_6 = 6 b_6^{\sigma}.
\end{aligned}
\ee
Using Eqs. (\ref{coeffs}) and (\ref{coeffs_shift}) one can see that all quadratic terms have strictly positive coefficients, whereas all quartic terms -- strictly negative coefficients. 

\subsection{Derivation of Eq. (\ref{deltaS_j})} \label{A:Integrate_out_charges}
Here we show that
\be
\begin{aligned}
	&\int d \tau_1 d \tau_2 dx_1 dx_2 f(x_1,\tau_2) G''_{\rho}(x_1-x_2,\tau_1-\tau_2) f(x_2,\tau_2) \approx -\frac{1}{2 B_2^{\rho}} \int d\tau dx f^2(x,\tau),\\
	&\int d \tau_1 d \tau_2 dx_1 dx_2 f(x_1,\tau_2) \ddot G_{\rho}(x_1-x_2,\tau_1-\tau_2) f(x_2,\tau_2) \approx -\frac{1}{2 A_2^{\rho}} \int d\tau dx f^2(x,\tau),
\end{aligned}
\ee
given that the Green's function satisfies
\be
	\Bigl( \partial_x^2 + (A_2^{\rho}/B_2^{\rho}) \partial_{\tau}^2 \Bigr)\, G_{\rho}(x,\tau) = - \frac{1}{B_2^{\rho}} \delta^2(x,\tau).
\ee
Going to the relative and the center of mass coordinates, we write the first integral as
\be
	\int dT dX dt dx \,  \partial_x^2 \,G_{\rho}(x,t) \, f( X+\frac{x}{2}, T+\frac{t}{2} ) f( X-\frac{x}{2}, T-\frac{t}{2} ) \approx \int dt dx G''_{\rho}(x,t) \int dT dX f^2(X,T).
\ee
Rescaling $y = \sqrt{B_2^{\rho}/ A_2^{\rho}} \tau$ brings the equation for $G_{\rho}(x,y)$ to
\be
	\Delta_{x,y} G_{\rho}(x,y) = -\frac{1}{ \sqrt{ A_2^{\rho} B_2^{\rho} } } \delta^2(x,y).
\ee
Since $G_{\rho}$ only depends on $r = \sqrt{x^2 + y^2}$, in polar coordinates we have
\be
	\left( \partial_r^2 + \frac{1}{r} \partial_r \right)G_{\rho}(r) = \Delta_{r} G_{\rho}(r) = -\frac{1}{ \sqrt{ A_2^{\rho} B_2^{\rho} } } \frac{\delta(r)}{2\pi r}.
\ee
Then, taking into account that
\be
	\partial_x^2 G_{\rho}(x,y) = \frac{1}{2}\Delta_r G_{\rho}(r) + \frac{\cos2\phi}{2} \left( \partial_r^2 - \frac{1}{r} \partial_r \right)
\ee
and the second term vanishes after the integration over the polar angle $\phi$, we write the integral $\int dx dt G_{\rho}''(x,t)$ in polar coordinates and get
\be
	 \int dt dx G''_{\rho}(x,t) = 2\pi \sqrt{\frac{A_2^{\rho}}{B_2^{\rho}}} \int_0^{+\infty} r dr \frac{1}{2}\Delta_r G_{\rho}(r) = -\frac{1}{2 B_2^{\rho} }.
\ee
which yields the first result stated in the beginning of the Appendix. The proof for the second one is identical.

\subsection{Effective spin Lagrangian in the $\Phi_{\sigma}$-representation}

Let us now rewrite the Lagrangian $\tilde{\cal L}_{\sigma}$ from Eq. (\ref{L_sigma_eff}) in the $\Phi$-representation. We begin by writing the corresponding Hamiltonian in the $\Theta$-representation (we omit the index $\sigma$, since we do not have the charge fields anymore):
\be \label{H_Theta}
	{\cal H}(\Pi_{\Theta}, \Theta') = \Bigl( \tilde{\cal L} + i \dot\Theta \Pi_{\Theta} \Bigr) \Bigr|_{\dot\Theta = \dot\Theta\left( \Pi_{\Theta} \right)} \,\, ,
\ee
where the expression for $\dot\Theta$ in terms of $\Pi_{\Theta}$ follows from
\be \label{Pi_Theta}
	i \Pi_{\Theta} = - \frac{\partial \tilde{\cal L}_{\sigma}}{\partial \dot\Theta} = -i a_1 - a_2 \dot \Theta -i a_3 \dot\Theta^2 + a_4 \dot\Theta^3 + i a_5 \dot\Theta^4 - a_6 \dot\Theta^5.
\ee
Since $\Pi_{\Theta} = - \Phi'$, one can also rewrite the above relation as 
\be \label{dtTheta_dxPhi}
	\dot\Theta = \frac{1}{a_2} \Bigl\{\,i\, \left( \Phi' - a_1 \right) -i a_3 \dot\Theta^2 + a_4 \dot\Theta^3 + i a_5 \dot\Theta^4 - a_6 \dot\Theta^5 \Bigr\}.
\ee
Then, using Eqs. (\ref{H_Theta}), (\ref{Pi_Theta}), and taking into account that $\Theta' = - \Pi_{\Phi}$, the $\Phi$-representation for the Hamiltonian can be formally written as
\be \label{H_formal}
	{\cal H} = \frac{b_2}{2} \Pi_{\Phi}^2 -\frac{b_4}{4}\Pi_{\Phi}^4 + \frac{b_6}{6} \Pi_{\Phi}^6 
	+ \Bigl(  -\frac{a_2}{2} \dot\Theta^2 
		- i \frac{2a_3}{3} \dot\Theta^3 
		+\frac{3a_4}{4}\dot\Theta^4
		+i\frac{4a_5}{5}\dot\Theta^5
		-\frac{5a_6}{6}\dot\Theta^6
		\Bigr)\Bigr|_{\dot\Theta = \dot\Theta\left( \Phi' \right)} \,\, ,
\ee
with $\dot\Theta$ being expressed via $\Phi'$ using Eq. (\ref{dtTheta_dxPhi}). 
Now, solving Eq. (\ref{dtTheta_dxPhi}) for $\dot\Theta$ in terms of $\Phi'$ by iterations, we get
\be \label{dtauTheta}
\begin{aligned}
	\dot\Theta &= \frac{i}{a_2}  \Phi' 
	+i \frac{a_3}{a_2^3} \Phi'{}^2 
	-i \frac{a_2a_4-2 a_3^2}{a_2^5} \Phi'{}^3
	+i\frac{5 a_3^3-5 a_2a_4 a_3+a_2^2 a_5}{a_2^7} \Phi'{}^4\\
	&+i \, \frac{ 14 a_3^4-21 a_2 a_4 a_3^2+6 a_2^2 a_5 a_3+a_2^2 \left(3 a_4^2-a_2a_6\right)}{a_2^9}\,\Phi'{}^5.
\end{aligned}
\ee
The potential part of Hamiltonian (\ref{H_formal}) becomes
\be
	\frac{1}{2 a_2}\Phi'{}^2
	+ \frac{a_3}{3 a_2^3} \Phi'{}^3
	+\frac{2 a_3^2-a_2 a_4}{4 a_2^5}\Phi'{}^4 
	+\frac{5 a_3^3-5 a_2 a_4 a_3+a_2^2 a_5}{5 a_2^7}{\Phi}'{}^5
	+\frac{14 a_3^4-21 a_2 a_4 a_3^2+6 a_2^2 a_5 a_3+3 a_2^2 a_4^2-a_2^3 a_6}{6 a_2^9} {\Phi }'{}^6
\ee
The Lagrangian in the $\Phi$-representation is then
\be \label{L_phi}
	\tilde{\cal L} ( \dot\Phi, \Phi' ) = \Bigl( {\cal H}\left( \Pi_{\Phi}, \Phi' \right) - i \dot{\Phi} \Pi_{\Phi} \Bigr) \Bigr|_{\Pi_{\Phi} = \Pi_{\Phi} (\dot{\Phi})} \,\, ,
\ee
where $\Pi_{\phi}$ is relates to $\dot{\Phi}$ via
\be
	i\dot{\Phi} = \frac{\partial {\cal H}}{\partial \Pi_{\Phi}} = b_2 \Pi_{\Phi} - b_4 \Pi_{\Phi}^3 + b_6 \Pi_{\Phi}^5.
\ee
The latter can be written as
\be
	\Pi_{\Phi} = \frac{1}{b_2} \left( i \dot{\Phi} + b_4 \Pi_{\Phi}^3 - b_6 \Pi_{\Phi}^5  \right).
\ee
Solving by iterations, we get
\be
	\Pi_{\Phi} = \frac{i}{b_2} \dot{\Phi} \left( 1 - \frac{b_4}{b_2^3} \dot{{\Phi }}^2 - \frac{ b_2 b_6-3 b_4^2}{b_2^6} \dot{{\Phi }}^4 \right).
\ee
The kinetic part of Hamiltonian (\ref{H_formal}) then gives
\be
	 \frac{b_2}{2} \Pi_{\Phi}^2 -\frac{b_4}{4}\Pi_{\Phi}^4 + \frac{b_6}{6} \Pi_{\Phi}^6  =  \frac{\dot{{\Phi }}^2}{2 b_2} -\frac{b_4}{4 b_2^4} \dot{{\Phi }}^4 + \frac{ 3 b_4^2-b_2 b_6 }{6 b_2^7} \dot{{\Phi }}^6.
\ee
Thus, using Eq. (\ref{L_phi}) we finally arrive at the Lagrangian given by (\ref{Lagr}) in the main text:
\be
\begin{aligned}
	\tilde{\cal L} ( \dot\Phi, \Phi' ) &= \frac{1}{2 b_2} \dot{{\Phi }}^2 +  \frac{1}{2 a_2} \Phi'{}^2
	+ \frac{a_3}{3 a_2^3} \Phi'{}^3
	+\frac{2 a_3^2-a_2 a_4}{4 a_2^5}\Phi'{}^4  \\
	&+\frac{5 a_3^3-5 a_2 a_4 a_3+a_2^2 a_5}{5 a_2^7}{\Phi}'{}^5
	+\frac{14 a_3^4-21 a_2 a_4 a_3^2+6 a_2^2 a_5 a_3+3 a_2^2 a_4^2-a_2^3 a_6}{6 a_2^9} {\Phi }'{}^6\\
	&\equiv \frac{A}{2} \dot\Phi^2 + \frac{u_2}{2}\Phi'^2 + \sum_{n=3}^6 u_n \Phi'{}^n.
\end{aligned}
\ee

\end{widetext}


\section{Momentum-shell RG} \label{AppMomShellRG}

In this Appendix we present a detailed derivation of the RG equations within the momentum-shell approach. We begin by considering the action in Eq. (\ref{GL_RG}) of the main text:
\be
\begin{aligned}
	&S[\Phi] = S_0[\Phi] + S_1[\Phi] \\
		&= \int dx d\tau \left\{ \frac{A}{2} \dot \Phi^2 + \frac{u_2}{2} \Phi'^2 \right\} + \int dx d\tau \sum_{n = 3}^6 u_n \Phi'^n
\end{aligned}
\ee
and expand the field into the slow and fast components as $\Phi = \Phi_< + \Phi_>$. The slow component $\Phi_<$ has momentum modes in the interval $0 < |k| < \Lambda/b$, whereas the fast component $\Phi_>$ --- in the interval $\Lambda/b < |k| < \Lambda$, where $\Lambda$ is the UV momentum cutoff and $b = \exp{(\delta l)}$ is the scaling factor. In the Gaussian part of the action the slow and fast components decouple: $S_0[\Phi] = S_{0<} + S_{0>}$. For the interaction part we have $S_1[\Phi_< + \Phi_>] = S_{1<} + \tilde S[\Phi_<, \Phi_>]$, where
\be
	\tilde S[\Phi_<, \Phi_>] = \int dx d\tau \sum_{n=3}^6 u_n \sum_{p=1}^n {n \choose p} \Phi'^{n-p}_< \Phi'^{p}_>.
\ee
Thus, the total action becomes $S[\Phi] = S_< + S_{0>} + \tilde S$. Then, expanding the partition function to second order in $\tilde S$, we obtain
\be
	{\cal Z} = \int {\cal D}\Phi_< e^{ - S_< - \langle \tilde S \rangle_{0>} + \frac{1}{2} \langle \tilde S^2 \rangle_{0>}^c },
\ee
where $\langle \tilde S^2 \rangle_{0>}^c = \langle \tilde S^2 \rangle_{0>} - \langle \tilde S \rangle_{0>}^2 $ and $\langle \ldots \rangle_{0>}$ is the average over the fast components $\Phi_>$ with the Gaussian action $S_{0>}$. In the above expression we omitted the constant contribution $\ln {\cal Z}_{0>}$.

\subsection{First order correction}

For the first order correction $\langle \tilde S \rangle_{0>}$ we need to calculate 
\be \label{first_order1}
	 \sum_{n=3}^6 \sum_{p=1}^n u_n { n \choose p} \Phi'^{n-p}_< \langle \Phi'^p_> \rangle_{0>}.
\ee
We immediately see that for $n=3,4$ we only need terms with $p=2$, and for $n=5,6$ only terms with $p=2,4$. Other terms either vanish upon averaging or give constant contribution independent of $\Phi'_<$. Let us first consider $p=2$ and calculate the corresponding Green's function $G_{0>} \equiv \langle \Phi'^2_> \rangle_{0>}$:
\be
	G_{0>} = \int_{-\infty}^{+\infty} \frac{d\omega}{2\pi} \int_> \frac{d k}{2\pi} \frac{k^2}{A\omega^2 + u_2 k^2}.
\ee
\linebreak
In the above expression we used a short-hand notation
\be
	\int_> \frac{dk}{2\pi} \equiv \int\displaylimits_{\Lambda/b < |k| < \Lambda} \frac{dk}{2\pi}.
\ee
Integrating over $d\omega$ and taking into account that $b\approx 1+ \delta l $, we get
\be \label{G1}
	G_{0>} = \frac{1}{2\pi} \frac{\Lambda^2}{(A  u_2)^{1/2}} \delta l \equiv {\cal G}_1 \delta l.
\ee
Consider now terms with $p = 4$. Using Wick's theorem one has $\langle \Phi'^4_> \rangle_{0>} = 3 G_{0>}^2$, which is $\sim \delta l^2$. Such terms do not contribute to the RG equations, derived in the limit $\delta l \to 0$. Therefore, for the first order correction $\langle \tilde S \rangle_{0>}$ we only need terms with $p=2$:
\be \label{first_order2}
\begin{aligned}
	\langle \tilde S \rangle_{0>} = {\cal G}_1 \delta l \int dx d\tau & \bigl\{ 3 u_3 \Phi'_< + 6 u_4 \Phi'^2_< + \\
	&+10 u_5 \Phi'^3_< + 15 u_6 \Phi'^4_< \bigr\}.
\end{aligned}
\ee
Note that a new term $\sim \Phi'_<$ has been generated. However, including it into the action does not lead to any new terms coming from $\langle \tilde S \rangle_{0>}$, as can be easily seen from Eq.~(\ref{first_order1}). We will see later that this is also true for $\langle \tilde S^2 \rangle_{0>}^c$.


\subsection{Second order correction}

We now turn to the calculation of the second order correction $\langle \tilde S^2 \rangle_{0>}^c$. Explicitly, it reads
\be \label{second_order1}
\begin{aligned}
	\langle \tilde S^2 \rangle_{0>}^c & = \int d1 d2 \sum_{n,m=3}^6 \sum_{p=1}^n \sum_{q=1}^m {n \choose p} {m \choose q} u_n u_m \\
	&\times \Phi'^{n-p}_<(1) \Phi'^{m-q}_<(2) \, \langle \Phi'^{p}_>(1) \Phi'^{q}_>(2) \rangle_{0>}^c,
\end{aligned}
\ee
where $d1d2 \equiv dx_1 d\tau_1 dx_2  d\tau_2$. Before doing any calculations, we note that applying Wick's theorem to $\langle \Phi'^{p}_>(1) \Phi'^{q}_>(2) \rangle_{0>}^c$ will in general produce a number of terms of the form 
\be \label{typical_term}
	\int d1 d2 \,  \langle \Phi'_>(1)\Phi'_>(2) \rangle_{0>}^{N} \, \Phi'^{n-p}_< (1) \, \Phi'^{m-q}_< (2),
\ee
where the exponent $N$ depends on $n$, $m$, $p$, and $q$. Obviously, not all such terms will contribute to the RG equations since we are only interested in those terms that are linear in $\delta l$. Therefore, in order to understand which terms we do need, let us first consider the case of arbitrary $N$. We then write
\be
	 \langle \Phi'_>(1)\Phi'_>(2) \rangle_{0>}^{N} = \prod_{j=1}^N \int_{\omega_j, k_j} \frac{k_j^2 e^{i k_j x -i\omega_j \tau} }{A \omega_j^2 + u_2 k_j^2},
\ee
where $x = x_1 - x_2$, $\tau = \tau_1 - \tau_2$, and 
\be
\int_{\omega_j, k_j} \equiv \int_{-\infty}^{+\infty}\frac{d\omega_j}{2\pi} \int_> \frac{dk_j}{2\pi}.
\ee
Since the momentum integral is over an infinitesimally small region, we put $k_j \approx \Lambda$ everywhere except for the exponent. A straightforward integration yields
\\
\\
\begin{widetext}
\be
	 \langle \Phi'_>(1)\Phi'_>(2)  \rangle_{0>}^{N} = \left( \frac{\Lambda}{2\pi} \right)^N \frac{1}{\left( A u_2 \right)^{N/2}} \, 
	 \exp{ \left\{ - \Lambda \, N \sqrt{\frac{u_2}{A} } \,|\tau| \right\} } \, 
	 \left( \frac{\sin \Lambda x - \sin \frac{\Lambda}{b}x}{x} \right)^N.
\ee

Then, using Picard representation of the delta function, $\lim_{M\to\infty} (M/2) \exp(-M |\tau| ) = \delta(\tau) $, we rewrite the above expression as
\be \label{2Pfun}
	\langle \Phi'_>(1)\Phi'_>(2)  \rangle_{0>}^{N} = \frac{1}{N \pi} \, \left( \frac{\Lambda}{2\pi} \right)^{N-1} \frac{1}{A^{(N-1)/2} \, u_2^{(N+1)/2}} \;\delta(\tau_1 - \tau_2) I^N(x_1 - x_2),
\ee
\end{widetext}
where we defined
\be
	I(x) =  \frac{\sin \Lambda x - \sin \Lambda(1-\delta l) x}{x}.
\ee

We now proceed by looking at the properties of $I^N(x)$. Consider an integral
\be \label{delta_fun_rep}
\begin{aligned}
	&\int_{-\infty}^{\infty} dx \left( \frac{\sin \Lambda x - \sin \Lambda(1-\delta l) x}{x} \right)^N f(x) \\
		&= \Lambda^{N-1}\int_{-\infty}^{\infty} \frac{dy}{y^N}\Bigl( \sin y - \sin(1-\delta l )y \Bigr)^N f\left( \frac{y}{\Lambda} \right)\\
		&\approx f(0) \Lambda^{N-1} \times 
			\begin{cases} 
				0 \,,  &\text{odd } N,\\
				\pi \, C_N \, \delta l^{N-1} \,, &\text {even } N.
			\end{cases}
\end{aligned}
\ee
In the above expression $C_N$ is a numerical coefficient ($0<C_N\leq 1$) and in the limit of large $\Lambda$ we approximated $f(y/\Lambda) \approx f(0)$. We see that essentially $I^N(x)$ is a representation of the delta function. In Eq. (\ref{2Pfun}) we then put 
\be \label{I^N}
	I^N(x_1 - x_2) \approx \pi C_N \Lambda^{N-1} \delta l^{N-1} \delta (x_1 - x_2).
\ee 

It follows immediately that for the calculation of $\langle \Phi'^p_>(1) \Phi'^q_>(2) \rangle_{0>}^c$ we only need $p$ and $q$ such that using Wick's theorem we get terms as in Eq. (\ref{typical_term}), but with $N = 2$. All other $N$ will give either zero or a contribution with higher powers of $\delta l$. The only way to get $N=2$ is by taking $p = q = 2$:
\be
\begin{aligned}
	\langle \Phi'^2_>(1)& \Phi'^2_>(2) \rangle_{0>}^c = 2 \, \langle \Phi'_>(1) \Phi'_>(2) \rangle_{0>}^2\\
	&= 2 \frac{ \Lambda^2 \delta l }{4\pi A^{1/2}  u_2^{3/2} } \delta(x_1 - x_2) \delta(\tau_1 - \tau_2),
\end{aligned}
\ee
where we used Eqs. (\ref{2Pfun}), (\ref{I^N}), and took into account that $C_2 = 1$.
Thus, the second order correction becomes
\be \label{second_order2}
\begin{aligned}
	\langle \tilde S^2 \rangle_{0>}^c  = 2 {\cal G}_2 \delta l \int dx d\tau  \sum_{n,m=3}^6 & {n \choose 2} {m \choose 2} u_n u_m \\
	&\times \Phi'^{n + m - 4}_<(x,\tau),
\end{aligned}
\ee
where we defined
\be \label{G2}
	{\cal G}_2 \equiv \frac{\Lambda^2}{4\pi \left(A\, u_2^3 \right)^{1/2} }.
\ee

Let us now recall that at the level of the first order correction the term $\sim \Phi'_<$ has been generated. We then mentioned that including such term into the action does not generate any additional terms under RG, even at the second order level. At this point it is easy to understand that this is indeed the case. Looking at Eq. (\ref{second_order1}) we see that, e.g., for $n=1$ there appear terms containing
\be
	\langle \Phi'_>(1) \Phi'^q_>(2) \rangle_{0>}^c.
\ee
Using Wick's theorem, one can make only one contraction between points 1 and 2, obtaining a factor of $\langle \Phi'_>(1) \Phi'_>(2) \rangle_{0>}$, which vanishes after the integration over $d2$ due to Eq. (\ref{delta_fun_rep}). For $m=1$ the reasoning is identical. 

Thus, all terms coming from the second order correction are already present in Eq. (\ref{second_order2}). Explicitly they are given by
\be \label{second_order3}
\begin{aligned}
	\langle \tilde S^2 \rangle_{0>}^c & = 2 {\cal G}_2 \delta l  \int dx d\tau \Bigl\{ 9 u_3^2 \Phi'^2_< + 36 u_3 u_4 \Phi'^3_< \\
	&+ (36 u_4^2 + 60 u_3 u_5) \Phi'^4_< \\
	&+ (120 u_4 u_5 + 90 u_3 u_6) \Phi'^5_< \\
	&+ (100 u_5^2 + 180 u_4 u_6) \Phi'^6_< \Bigr\},
\end{aligned}
\ee
where we did not include newly generated terms $300 u_5 u_6 \Phi'^7_<$ and $225 u_6^2 \Phi'^8_<$ since they are beyond the initial expansion order of the GL functional  (\ref{GL_RG}).


\subsection{Renormalized action and RG equations}

We write the renormalized action as
\be
\begin{aligned}
	&S[\Phi_<] = \int dx d\tau \Bigl\{  \frac{A}{2} \dot \Phi_<^2 + \frac{u_2 + \delta u_2 \delta l}{2} \Phi_<'^2 \\
		& + (u_1 + \delta u_1 \delta l) \Phi'_< + \sum_{n=3}^6 (u_n + \delta u_n \delta l) \Phi'^n_< \Bigr\},
\end{aligned}
\ee
where $\delta u_j$ follow from Eqs. (\ref{first_order2}) and (\ref{second_order3}):
\be
\begin{aligned}
	\delta u_1 &= 3 {\cal G}_1 u_3, \\
	\delta u_2 &= 12 {\cal G}_1 u_4 - 18 {\cal G}_2 u_3^2, \\
	\delta u_3 &=  10 {\cal G}_1 u_5 -  36 {\cal G}_2 u_3 u_4 , \\
	\delta u_4 &= 15 {\cal G}_1 u_6 - 36 {\cal G}_2 u_4^2 -  60 {\cal G}_2 u_3 u_5, \\
	\delta u_5 &= - 120  {\cal G}_2 u_4 u_5 - 90 {\cal G}_2 u_3 u_6, \\
	\delta u_6 &= - 100  {\cal G}_2 u_5^2 - 180 {\cal G}_2 u_4 u_6.
\end{aligned}
\ee
The quantities ${\cal G}_1$ and ${\cal G}_2$ are defined in Eqs. (\ref{G1}) and (\ref{G2}), correspondingly.
Making the rescaling $x = b \tilde x$, $\tau = b^z \tilde \tau$, and $\Phi_<(x,\tau) = b^{\chi} \tilde \Phi_< (\tilde x, \tilde \tau)$ we obtain the RG equations:
\be
\begin{aligned}
	A(b) &= b^{2\chi - z +1}A,\\
	u_n(b) & = b^{n\chi + z - n + 1}(u_n + \delta u_n \delta l), \quad n = 1,\ldots,6.
\end{aligned}
\ee
Taking $b\approx 1 + \delta l$, in the limit $\delta l \to 0$ we obtain the RG equations in the differential form:
\be \label {diffRGeqs}
\begin{aligned}
	\partial_l A &= (2\chi -z + 1)A, \\
	\partial_l u_1 &= (\chi+z)u_1 + 3 \,{\cal G}_1 \, u_3,\\
	\partial_l u_2 &= (2\chi + z-1) u_2 + 12 \, {\cal G}_1 \, u_4 - 18 \, {\cal G}_2 \, u_3^2,\\
	\partial_l u_3 &= (3\chi+z-2) u_3 + 10 \, {\cal G}_1 \, u_5 -  36 \, {\cal G}_2 u_3 u_4,\\
	\partial_l u_4 &= (4\chi+z-3)u_4 + 15 \, {\cal G}_1 \, u_6 - 36 \, {\cal G}_2 \, u_4^2 -  60 \, {\cal G}_2 \, u_3 u_5,\\
	\partial_l u_5 &= (5\chi+z-4) u_5 - 120 \, {\cal G}_2 \, u_4 u_5 - 90 \, {\cal G}_2 \, u_3 u_6, \\
	\partial_l u_6 &= (6\chi+z-5) u_6 - 100 \,  {\cal G}_2 \, u_5^2 - 180 \, {\cal G}_2 \, u_4 u_6.\\ 
\end{aligned}
\ee


\begin{thebibliography}{99}
\bibitem{40K}
	A remarkable candidate for experimental verification is the gas of ultracold $^{40}$K atoms, where in three dimensions the $s$- and $p$-wave Feshbach resonances are very close to each other. Thus, one can have $s$-wave repulsion and $p$-wave attraction that are both resonant.
\bibitem{instability}
For $y > 2/7$ the bare coupling $g_{6}(0)$ becomes negative and our theory becomes thermodynamically unstable. In this region, in order to stabilize the theory, one has to take into account even higher order terms in the Lagrangian $(\ref{Lagr})$.
\bibitem{fixed_points}
Apart from the Gaussian fixed point, the system of RG equations (\ref{RG_g}) possesses a number of non-trivial fixed points. However, a detailed analysis shows that in our case none of them are reachable if one starts from realistic initial conditions.
\end{thebibliography}
\end{document}